\documentclass[12pt]{JHEP3}
\usepackage{epsfig}
\preprint{ {\tt hep-th/0311086} }
\newcommand{\be}[1]{ \begin{equation}\label{#1} }
\newcommand{\ee}{\end{equation}}
\newcommand{\bea}[1]{\begin{eqnarray}\label{#1} }
\newcommand{\eea}{\end{eqnarray}}
\newcommand{\eq}[1]{(\ref{#1})}

\title{Partial ${\cal N}=2 \rightarrow {\cal N}=1$ supersymmetry breaking and gravity deformed chiral rings}
\author{ Justin R. David$^a$, Edi Gava$^{a,b}$ , K. S. Narain$^a$ \\
$^a$High Energy Section, \\
The Abdus Salam International Centre for Theoretical Physics,
\\Strada Costiera, 11-34014 Trieste, Italy.\\
$^b$Instituto Nazionale di Fisica Nucleare, sez. di Trieste, \\
and SISSA, Italy. \\
\email{justin, gava, narain@ictp.trieste.it} 
}
\abstract{
We present a derivation of the chiral ring relations,
arising in  ${\cal N}=1$ gauge theories in the presence of 
(anti-)self-dual
background gravitational field $G_{\alpha\beta\gamma}$ and 
graviphoton field strength $F_{\alpha\beta}$. These were previously
considered in the literature in order 
to prove the relation between gravitational
F-terms in the gauge theory and coefficients of the topological
expansion of the related matrix integral. We consider 
the spontaneous breaking of ${\cal N} =2$ to ${\cal N} =1$  
supergravity coupled to vector- and hyper-multiplets, 
and take a rigid limit which keeps a non-trivial  $G_{\alpha\beta\gamma}$    
and $F_{\alpha\beta}$ with a finite supersymmetry breaking
scale. We derive the resulting effective, global, ${\cal N}=1$ theory
and show that the chiral ring relations are just a consequence of
the standard ${\cal N}=2$ supergravity Bianchi identities . We can also
obtain models with matter in different representations and in 
particular quiver theories.
We also show that, in the presence of non-trivial
$F_{\alpha\beta}$, consistency of the Konishi-anomaly loop equations with
the chiral ring relations, demands that the gauge kinetic
function and the superpotential, 
a priori unrelated for an ${\cal N}=1$ theory, should
be derived from a prepotential, indicating an underlying
${\cal N}=2$ structure.
}

\begin{document}
\baselineskip 4ex

\section{Introduction}

In the last year there has been a considerable progress in 
computing glueball superpotentials in ${\cal N}=1$ Super Yang-Mills
(SYM) theories,
as a result of the discovery made in \cite{Dijkgraaf:2002dh} 
that such computations
can be efficiently organized by a matrix model integral. 
Whereas the leading planar term of the latter is related to ordinary
F-terms in the gauge theory, higher genus contributions in the
topological expansion of the matrix model
capture certain generalized gravitational F-terms, originally
studied in the context of type II string theories on Calabi-Yau 
manifolds \cite{Antoniadis:1994ze,Bershadsky:1994cx}.
In particular, the genus 1 term computes
an F-term involving the gravitational background $G_{\alpha\beta\gamma}$
to which the gauge theory couples. The non-trivial gravitational background    
is responsible for a deformation of the basic chiral ring relation
among the gauginos of the ${\cal N}=1$ SYM theory, 
as can be seen by starting from the Bianchi
identities of ${\cal N}=1$ supergravity coupled to super Yang-Mills.
This fact plays a crucial role in the proof of the identification
of the superpotential term proportional to $G^2$
with the leading non-planar term in the matrix model integral.

As for the $g>1$ terms however, the generalized superpotentials
necessarily involve
the graviphoton field strength $F_{\alpha\beta}$
and therefore cannot be understood 
within the framework of ${\cal N}=1$ supergravity and the related
tensor calculus. In \cite{Ooguri:2003qp,Ooguri:2003tt} 
it has been proposed, with
string theoretic motivations, that the basic chiral ring relation,
stating the Grassmann nature of the gluino superfield $W_\alpha$, 
is deformed
by the presence of a non-trivial graviphoton field strength, whereas 
superspace coordinates are unaffected. A different
viewpoint has been taken in \cite{deBoer:2003dn,Seiberg:2003yz}, 
where instead the
Grassmann nature of the superspace coordinates is deformed.
 
The aim of this paper is to reconsider the above issue. 
In \cite{Alday:2003ms} we have shown that the chiral ring relations 
one obtains for gauge invariant fields from the deformed chiral ring relation
$\{W_\alpha, W_\beta\}=F_{\alpha\beta}+
2G_{\alpha\beta\gamma}W^\gamma$, can be obtained from the Bianchi
identities of ${\cal N}=2$ supergravity coupled to abelian
vector multiplets and spontaneously broken by fluxes, in accordance
with the open string (gauge theory)/closed string duality.
 
We would like to follow the same strategy on the 
super Yang-Mills side (we will discuss also the possibility
that this picture may arise from wrapped D5-branes effective action).
The idea here is to start from ${\cal N}=2$ supergravity coupled to 
vector multiplets in the adjoint representation of a gauge
group, that will be taken to be $U(N)$ (we will consider afterwards 
also hypermultiplets in the fundamental of $U(N)$), break it spontaneously
to ${\cal N}=1$, following the procedure in
\cite{Ferrara:1996gu,Ferrara:1996xi} of gauging
two commuting translation isometries of the
hypermultiplet quaternionic manifold $SO(4,1)/SO(4)$, and then take
a scaling limit that decouples the quantum gravitational sector
and hypermultiplet sector,
while keeping a non trivial background for $F_{\alpha\beta}$
and $G_{\alpha\beta\gamma}$. This involves taking the Planck mass
$m_P$ to infinity, with finite supersymmetry
braking scale $\Lambda$ and scaling consistently
the various fields. The resulting theory is an effective
${\cal N}=1$ $U(N)$ gauge theory with a non trivial superpotential
for the adjoint chiral superfield $\Phi$, coupled to the
gravitational background of $F$ and $G$. Then, starting from the 
Bianchi identities of ${\cal N}=2$
supergravity \cite{Andrianopoli:1997cm} 
and expanding around the chosen vacuum we will 
derive precisely the chiral ring relation mentioned    
in the previous paragraph. However, before claiming that we have obtained
the super Yang-Mills  
theory considered in \cite{Dijkgraaf:2002dh}, we have to settle two
issues: the first is that, since our ${\cal N}=1$ SYM theory comes from a
spontaneously broken  ${\cal N}=2$ theory, a non-trivial superpotential for
$\Phi$ implies also a non-trivial, $\Phi$-dependent gauge function, which
is related to the superpotential in a specific way, since both
are obtained from a prepotential.
The second problem is that the ${\cal N}=2$ Bianchi identities that give
the wanted chiral ring relation, also give a relation of the form
$[W_\alpha, \Phi] \sim F_{\alpha\beta}W^\beta$, i.e. $W$ and $\Phi$ do not
commute in the ring. 

Quite remarkably, the two problems cure each other:
$\Phi$ can be redefined by an appropriate shift 
proportional to $W^2$ in such a way
that it commutes with $W$. Then, after performing the same shift   
in the effective action, i.e. in the superpotential and in the
gauge function, one can show that all the non-trivial $\Phi$ and $\Lambda$
dependence in the latter cancels in the chiral ring, 
therefore leaving us with precisely
the SYM theory considered in \cite{Dijkgraaf:2002dh}. 

We should stress here an important result of
our analysis: in presence of a non-trivial $F_{\alpha\beta}$, 
consistency of the loop equations with the chiral
ring relations, demands that the gauge function,
a priori arbitrary for an ${\cal N}=1$  super Yang-Mills theory,
must be the one given by ${\cal N}=2$ supersymmetry. 
We regard this fact as a strong indication that 
our interpretation in terms of spontaneously broken
${\cal N}=2$ supergravity is the correct one.

Finally, of course, the underlying ${\cal N}=2$
structure also implies that the adjoint scalar superfield comes
with a non-trivial kinetic K\"{a}hler potential, which gives
rise to additional non-renormalizable interactions. However, from
the point of view of  ${\cal N}=1$ super Yang-Mills, this is a D-term and
therefore is not expected to modify the loop equations
in the chiral ring coming from the generalized Konishi anomaly.

The above strategy can be extended to include the case of matter in
the (anti-)fundamental representation of the gauge group 
\cite{Seiberg:2002jq}. To cover
this case one has to add to the previous case hypermultiplets. We start
with hypermultiplets parametrizing the quaternionic manifold
$U(2,N+1)/U(2)\times U(N+1)$ and vector multiplets in the $U(1)\times
U(N)$ and gauge appropriately the hypermultiplets \cite{Fre:1997js} 
to achieve supersymmetry breaking and $U(N)$-charged fundamental
hypermultiplets. In this way one gets, after decoupling, 
an effective ${\cal N}=1$ super Yang-Mills 
theory with a trilinear coupling of $\Phi$
to chiral superfields in the $\bf N$ and $\bar{\bf N}$. Again, considering
the Bianchi identities involving hypermultiplets, one can show
that the chiral ring relations involving fundamentals are unmodified
by the non trivial gravitational background. Other gauge groups 
and matter representations can also be incorporated in the discussion.
In particular, quiver theories can also be obtained in this way.   

An obvious question finally arises, as to whether the above
picture can be given a D-brane interpretation. In the
closed/open string duality of \cite{Vafa:2000wi}, we are dealing, in
the open string side, with D5-branes wrapped on two-cycles
inside certain non-compact Calabi-Yau manifolds. We have seen, in
the field theoretic discussion, that what triggers partial
supersymmetry breaking is the gauging of two translation isometries
in the hypermultiplet's manifold. That is, more 
precisely, certain scalars in hypermultiplets are charged under 
the graviphoton $U(1)$ and the ``center of mass'' 
$U(1)$ in $U(N)$. Whereas the former is a bulk
field, it is natural to identify the latter with
the center of mass degree of freedom of the D5-branes.   
In fact one can show that there is the correct coupling of
the center of mass $U(1)$ with hypermultiplets from the closed
string sector (this was also noticed in
\cite{Douglas:1996sw}). However, due to the non-compactness of the
Calabi-Yau manifold, the issue of whether there is a gauging by the
graviphoton is not clear.    

The paper is organized as follows:
in section 2 we review the gauging formalism in ${\cal N}=2$
supergravity coupled to vector- and hyper-multiplets and discuss
the gauging, which gives rise to partial supersymmetry breaking,
in the case where only $U(N)$ vector multiplets are charged, after
taking the rigid limit $m_P\rightarrow \infty$. In section 3
we also explicitly
write down the resulting ${\cal N}=1$ effective action.
In section 4 we include hypermultiplets 
valued in the $U(2,N+1)/U(2)\times U(N+1)$ quaternionic manifold,
discuss their gauging, and the resulting
supersymmetry breaking pattern. We show that in the rigid limit 
we obtain the coupling of $\bf{ N}$ and $\bar{\bf{N}}$ chiral multiplets
to the adjoint multiplet $\Phi$ and to the gauge multiplet $W$.
In section 5 we derive the chiral ring relations from the solution
of the Bianchi identities of  ${\cal N}=2$ supergravity of 
\cite{Andrianopoli:1997cm} 
after expanding around the vacuum which partially
breaks supersymmetry. In section 6 we discuss the generalized Konishi
anomaly equations and show that, in the presence of 
$F_{\alpha\beta}$, consistency demands that the gauge function 
and the superpotential should both be derived from a prepotential,
indicating an underlying ${\cal N}=2$ structure.

In appendix A we explain the conventions and notation
we follow for the hypermultiplets, while in appendix B we
prove that after the field redefinition for $\Phi$ the
gauge function become trivial in the chiral ring.

\section{${\cal N} =2$ gravity, gaugings and the rigid limit }

In this section we 
provide the details of the ${\cal N}=2$ supergravity 
coupled with $U(N)$ vector multiplets, the gaugings of the vector
multiplet manifold and
the scalings  of the fields with the Planck mass
to obtain ${\cal N}=1$ theory in the rigid limit. 
The details of the ${\cal N}=2$ supergravity action along with the 
gaugings of the matter fields were developed in a series of works
by various authors \cite{ deWit:1980ug,deRoo:1980mm,Bergshoeff:1981is,
Cremmer:1985hj,deWit:1985px}, for
the purposes of this paper 
we use the Lagrangian and the conventions of \cite{Andrianopoli:1997cm}. 
Partial breaking of ${\cal N}=2$ to ${\cal N}=1$ requires gauging a
minimal hypermultiplet manifold by the $U(1)$ corresponding to the
graviphoton and the overall $U(1)$ \cite{Ferrara:1996gu}.
For concreteness we take the minimal hypermultiplet manifold to be
$SO(4,1)/SO(4)$. The theory we discuss below is the generalization of
the one studied by \cite{Ferrara:1996xi} for the 
case of a $U(1)$ vector multiplet. 
We obtain the rigid limit by scaling the Planck mass to
infinity in such a way so as to retain a non dynamical supergravity
background. This method provides a natural way to derive
the couplings of the background ${\cal N} =2$ supergravity fields to the 
matter of the ${\cal N} =1$ theory after spontaneous breaking of
supersymmetry.

\subsection{Field content}

${\cal N}=2$ supergravity consists of the gravity multiplet and the 
matter multiplets. The gravity multiplet contains the graviton, two
gravitinos $\psi_{\mu}^A$, with $A=1,2$  
and a $U(1)$ gauge field called the
graviphoton. The matter multiplets are of two types, the vector
multiplets and the hypermultiplets. We provide the details of this
sector below. 

\vspace{1ex}
\noindent \emph{Vector multiplets}

The scalars of the  $U(N)$ 
vector multiplets parametrize a  $2N^2$ dimensional 
special K\"{a}hler manifold which
is specified by following holomorphic section.
\bea{vecsec}
X^{0}(z)  = \frac{1}{\sqrt{2}}, \;\;&\;&\;\;\;  
F_{0}(z) = -\frac{i}{\sqrt{2} }
\left( 2 f(z) - z^a \frac{\partial f( z) }{\partial z^a} \right) , \\
\nonumber
X^i(z) =   \frac{z^i}{\sqrt{2}} 
\;\;&\;&\;\;\;  F_i (z) = 
- \frac{i}{\sqrt{2}} \frac{\partial f(z) }{\partial z^i}, 
 \\ \nonumber
X^n(z) = \frac{i}{\sqrt{2} } \frac{\partial f(z) }{\partial z^n}, 
\;\;&\;&\;\;\;
F_n(z) = \frac{z^n }{\sqrt{2}}. 
\eea
Here the $z$'s are complex coordinates with indices
$i, j$ which take values from $\;$
$1, \ldots , (n-1)$, while $a, b$ take
values from $1, \ldots, n$  and $n= N^2$, the number of generators of
$U(N)$. The $i, j$ indices parametrize the $SU(N)$ directions while
the $n$ index refers to the overall $U(1)$ of $U(N)$.
We normalize the Killing metric of $SU(N)$ such that
$\rm{Tr} ( T_i T_j)= \delta_{ij}$, 
where $T_i$ stands for $N\times N$ Hermitian matrices which
generate $SU(N)$.   
The last holomorphic section in \eq{vecsec} is
necessary to make the structure group of the ${\cal N }= 2$ theory
$U(N)$.  The action of the group $U(N)$ on the coordinates $z^a$ is
given by
\be{acgroup}
U z U^\dagger , \;\;\;\;\; \hbox{ where } \;\;z = 
( z^iT_i + z^n \frac{I}{\sqrt{N}}   ) ,
\ee
where $I$ is the identity matrix.
$f(z)$ is any function which is invariant under the action of
$U(N)$ on the coordinates $z$. 
${\cal N} =2$ supersymmetry restricts the  scalar manifold to
be special K\"{a}hler, therefore these  sections must satisfy
\be{skcond}
X^{\Lambda} \partial_a F_\Lambda - 
F_\Lambda \partial_a X^\Lambda =0
\ee
here  $\Lambda = 0, \ldots n$. 
It is easy to verify  from the ans\"{a}tz for the holomorphic symplectic
sections given in \eq{vecsec}, the above condition is obeyed. 
The K\"{a}hler potential is given by
\bea{kahpot}
K &=& - \log i ( X^{*\Lambda} F_\Lambda - X^\Lambda F^*_\Lambda ), \\
\nonumber
&=& - \log [  ( f + f^* ) - 
\frac{1}{2}( z^a - z^{a *}  ) ( \partial_a f - ( \partial_a f)^*)  ],
\eea
from which the the metric on the manifold is constructed  by $g_{ab*} =
\partial_a\partial_{b*} K$. 

We now discuss the gauging of the vector multiplet manifold. To
perform the gauging we need to construct the Killing vectors of the
manifold.  From the construction of the symplectic sections in
\eq{vecsec} and the adjoint action of $U(N)$ on the
coordinates $z^a$ given in \eq{acgroup}, 
we see that the following are the $N^2 -1$ killing
vectors corresponding to the $SU(N)$ subgroup of $U(N)$.
\bea{kill}
 k_i^k \partial_k = f_{ij}^k z^j \partial_k , \;\;&\;&\;\; 
 k_i^{k*} \partial_{k*} = f_{ij}^{k} z^{j*} \partial_{k*}, \\ \nonumber
{\bf k}_i  &=& k_i^k \partial_k + k_i^{k*} \partial_{k*}
\eea
Here $f_{ab}^c $ are the structure constants satisfying
\be{strcon}
[T_i, T_j] = i f_{ij}^kT_k,
\ee
note that the killing vector satisfies the following commutation
relation
\be{kilcom}
[{\bf k}_i , {\bf k}_j ] = -f_{ij}^k {\bf k}_k.
\ee
${\cal N} =2$ supersymmetry requires that the symplectic sections transform in
the adjoint representation of $U(N)$, this is seen from the following
equations
\bea{adjsec}
k^k_i \partial_k X^j = f_{ik}^j X^k, \;\;\;\;\; 
k^k_i \partial_k F_j = - f_{ij}^k F_k
\eea
These equation ensures that the following constraints are
satisfied
\be{kconstr}
k^j_i X^i = k^{j*}_{i*} {X}^{*i} = 0
\ee
The prepotential functions which generate these
Killing vectors
\be{kprepot}
{\cal P}_i = -i k^j_i \partial_j K
\ee
Given this definition of the prepotential functions, and using \eq{kilcom}
and the fact that the K\"{a}hler two form is closed, one can show
\be{comiden}
g_{ij*}( k_k^i k_l^{j*} - k_l^i k_k^{j*} ) = f^i_{kl} {\cal P}_i
\ee
This concludes our discussion of the vector multiplets. The above
informations are sufficient to write down the explicit Lagrangian,
which is  given in \cite{Andrianopoli:1997cm}.

\vspace{1ex}

\noindent\emph{Hyper multiplets }

For partial breaking of ${\cal N}=2$ supersymmetry it is necessary that 
that  that the vectors are coupled to at least one hypermultiplet
\cite{Ferrara:1996gu}.
The reason for this is that when ${\cal N}=2$ is broken to ${\cal
N}=1$ the massive gravitino is part of a massive spin 3/2  
multiplet which contains two massive spin one fields. Thus the
$U(1)$, corresponding to the graviphoton, and the overall $U(1)$ should
become massive by  Higgs mechanism. As the scalars of the 
vectors are in the adjoint representation, there must be  
at least two scalars from the hypermultiplet sector which are eaten up by two
$U(1)$'s.  Thus we have the condition that there must be at least one
hypermultiplet and at least two $U(1)$ translational isometries of the
hypermultiplet manifold which provide the Higgs mechanism. 
To be specific about the hypermultiplet manifold, 
we choose to work with the minimal model, which has   
a single hypermultiplet parametrizing the coset $SO(4,1)/SO(4)$
\footnote{In the next section we will generalize our discussion to
the manifold $U(M, 2)/U(M)\times U(2)$}
using coordinates $b^u, u = 0, 1, 2, 3$. 
As $SO(4,1)/SO(4)$ is Euclidean $AdS_4$ the metric on this
quaternionic manifold can be chosen to b $h_{uv} = (\frac{1}{b^0})^2
\delta_{uv}$. 
We choose  the $SU(2)$
connection $\omega^x $ on the quaternionic manifold as
\be{sucon}
\omega^x = \omega^x_u db^u , \;\;\;\;\;\; \omega_u^x = \frac{1}{b^0}
\delta^x_u 
\ee
where $x = 1, 2, 3$. 
The quaternionic potentials is given by the field strength of the
$SU(2)$ connections, they are given by
\bea{qp}
\Omega^x = \Omega^x_{uv} db^u\wedge db^v = d\omega^x + \frac{1}{2}
\epsilon^{xyz} \omega^y\omega^z, \\ \nonumber
\Omega^x_{0u} = -\frac{1}{2(b^0)^{2} } \delta^x_u, \;\;\;\;\;
\Omega^x_{yz} = \frac{1}{2(b^0)^2 } \epsilon^{xyz}
\eea

To gauge the hypermultiplet manifold we need to introduce the
corresponding Killing vectors. We choose to gauge the hypermultiplet
manifold by two translations corresponding to the $U(1)$ of the
graviphoton and the overall $U(1)$ of the structure group $U(N)$. 
The Killing vectors  and the  triplet of
prepotentials ${\cal P}_\Lambda^x$ generating them, are given as follows
\bea{hygaug}
k_0^u = g_1 \delta^{u1} + g_2 \delta^{u2}, \;\;\;\;\;
k_n^u = g_3 \delta^{u2}, \\ \nonumber
{\cal P}_0^x = \frac{1}{b^0} ( g_1 \delta^{x1} + g_2 \delta^{x2} ),
\;\;\;\; {\cal P}_n^x = g_3 \frac{1}{b^0} \delta^{x2}, \;\;\; x=1,2,3
\eea
It is easy to check that the prepotentials and the Killing vectors
satisfy the Poisson bracket condition for abelian gauge groups, given by
\be{pba}
\Omega_{uv}^x k_0^uk_n^v - \frac{1}{2} \epsilon^{xyz} {\cal
P}_0^y{\cal P}_n^z =0
\ee

To summarize,  we have the following field content, ${\cal N} =2$
gravity multiplet,
${\cal N}=2$ vector multiplet gauged under the group $U(N)$
and a hypermuliplet charged under the $U(1)$ corresponding to the
graviphoton and the overall $U(1)$ of $U(N)$. For future reference 
we write down the 
general form for the scalar potential which arises due to the 
gaugings of the vector multiplet manifold and the hypermultiplet
manifold
\bea{fulscalpot}
V(z, \bar{z}, b) &=& 
g^2\left[ 
\left( g_{ab*} k^a_\Lambda k^{b*}_\Sigma + 
4 h_{uv} k^u_\Lambda k^v_\Sigma \right) \bar{L}^\Lambda L^\Sigma 
\right. \\ \nonumber
&+& \left. 
g_{ab^*} f^\Lambda_a f^\Sigma_{b*} {\cal P}^x_\Lambda {\cal P}^x_\Sigma
- 3 \bar{L}^\Lambda L^\Sigma {\cal P}^x_\Lambda {\cal P}^x_\Sigma
\right]
\eea
Here $L^\Lambda = e^{K/2} X^\Lambda$ and 
$f^\Lambda_a = (\partial_a + \frac{1}{2} \partial_a K ) L^\Lambda$. Note that 
every term in the scalar potential depends either on the 
moment maps or the Killing vectors.

\subsection{Planck mass scalings}

To obtain the rigid limit we need to take the Planck mass $m_p$ to infinity. 
In this section we detail the Planck mass scalings of all the quantities 
involved in the ${\cal N}=2$ action, so that we can obtain the effective
field theory in rigid limit. 
The fields and the coordinates of the 
${\cal N}=2$ supergravity action as written in 
\cite{Andrianopoli:1997cm} are all dimensionless.
To restore the canonical dimensions of the coordinates they are 
scaled as $x^\mu \rightarrow m_p x^\mu$, while the supersymmetry parameter
or the superspace coordinate is scaled as $\epsilon^A  \rightarrow
\sqrt m_p \epsilon^A$. 
We follow  a similar procedure  for all
the quantities in the Lagrangian. 
First we discuss the scalings involved  in  the geometry, for the
function $f$ in \eq{vecsec},  following \cite{Ferrara:1996xi} 
we assume the following scaling form
\be{scalef}
f(z) = \frac{1}{2} + \frac{\Lambda}{m_p} z^n + \frac{\Lambda^2}{m_p^2}
\phi(z)
\ee
here the scale $\Lambda$ will be shown to be  the supersymmetry breaking scale subsequently,
$\phi$ is an $U(N)$ invariant function of $z^i$. Note that the linear
term is a function only of $z^n$.  The sections $X^i$ and $F_i$ are
scaled as
\be{secscale}
X^i \rightarrow \frac{\Lambda}{m_p} X^i, \;\;\;\;\;\;
F_i \rightarrow \frac{m_p}{\Lambda} F_i
\ee
With these scalings for the holomorphic sections and 
$f$ given by \eq{scalef} we write down  the
scaling form  of the following geometric quantities, 
which will be repeatedly used in our analysis
\bea{scaleqan}
\partial_i K &=& -\frac{\Lambda^2}{m_p^2} \left(
\frac{1}{2} (\partial_i \phi + \partial_{i^*} \phi^* )
-\frac{1}{2} ( z^a - z^a) \partial_i\partial_a \phi \right) + O
(\frac{\Lambda^3}{m_p^2})  , \\
\nonumber
\partial_n  K &=& -\frac{\Lambda}{m_p} + O (\frac{\Lambda^2}{m_p^2} ), \\
\nonumber
g_{ij^*} &=& - \frac{\Lambda^2}{m_p^2} \frac{1}{2}
\left( \partial_i\partial_j \phi + \partial_{i*}\partial_{j*} \phi^*
\right) + O (\frac{\Lambda^3}{m_p^3} ) , \\ \nonumber
g_{in^*} &=& - \frac{\Lambda^2}{m_p^2} \frac{1}{2}
\left( \partial_i\partial_n \phi + \partial_{i*}\partial_{n*} \phi^*
\right) + O (\frac{\Lambda^3}{m_p^3} ) , \\ \nonumber
g_{nn^*} &=& \frac{\Lambda^2}{m_p^2} \left( 1 - \frac{1}{2}
\partial_n\partial_n \phi - \frac{1}{2} \partial_{n*}\partial_{n^*}
\phi^* \right) + O (\frac{\Lambda^3}{m_p^3} ) , \\ \nonumber
\eea
Note that the scaling behaviour of $\partial_i K$ and the
K\"{a}hler metric is consistent with equation \eq{comiden} 
relating the Killing vectors to the prepotential.
The couplings of the hypermultiplet to the $U(1)$'s are scaled as
follows
\be{scalchy}
g_1 = \frac{\Lambda^2}{m_p^2} \xi, \;\;\;\;\; g_2 =
\frac{\Lambda^2}{m_p^2} e, \;\;\;\;\; g_3 = 2\frac{\Lambda}{m_p} m 
\ee

We now go over to discuss the scalings of the fields. 
We scale the ${\cal N}=2$ gravity fields so as to retain a
non-dynamical gravity background while the  gravity fluctuations are scaled
so that they decouple from the matter sector in the $m_p \rightarrow
\infty$ limit. Thus the resulting theory has only global
supersymmetry, but it is coupled to the ${\cal N}=2$ gravity
backgrounds.
The fields of the gravity multiplet scale as
\bea{gfscale}
V_\mu^a \rightarrow V_\mu^{a(b)} + \frac{1}{m_p} \delta V_\mu^a , \;\;\;\;\;
T_{\mu\nu} \rightarrow \frac{1}{m_p} T_{\mu\nu}^{(b)} + 
\frac{1}{m_p^2} \delta T_{\mu\nu} , \\ \nonumber
\Psi_{A\mu} \rightarrow  \frac{1}{\sqrt{m_p}} \Psi_{A\mu}^b + 
\frac{1}{m_p^{3/2}} \delta\Psi_{A\mu} . 
\eea
Here $V_\mu^a$ stands for the vielbein, $T_{\mu\nu}$ refers to the
graviphoton field strength and $\Psi_{A\mu}$ refer to the two
gravitinos of the ${\cal N}=2$ gravity multiplet.
Note that the ratio of the fluctuations to the background is
suppressed
by $1/m_p$. As a simple example  for the decoupling of the matter sector
from the fluctuations of the supergravity field,  consider a
minimally coupled scalar whose kinetic term is canonically normalized.
It is easy to see  then that the 
interaction with the fluctuation of the metric is down by  a
factor of $1/m_p^2$.   

In the matter sector, the vector multiplets scale as
\be{vscale}
G_{\mu\nu}^a \rightarrow \frac{1}{\Lambda m_p} G_{\mu\nu}^a,
\;\;\;\;\;\;
\lambda_A^a \rightarrow \frac{1}{\Lambda \sqrt{m_p} } \lambda_A^a
\ee
where $G^a_{\mu\nu}$ is the gauge field strength and $\lambda^a_A$
are the two gauginos. 
These scalings ensure that the kinetic terms for the gauginos
and for the gauge fields are canonically normalized. 
We do have to scale also the scalars of the vector multiplets by
$z \rightarrow  z/\Lambda$, in order to canonically normalize their
kinetic term, however for convenience we will reserve this for section
2.4. However there are terms in the Lagrangian  which couple the 
gravitational backgrounds of \eq{gfscale}  to the matter sector and
survive the $m_p\rightarrow \infty$ limit. Though we will not keep
track of these terms explicitly, we will incorporate them in the
analysis of chiral ring equations in section 5.

Finally the hypermultiplets scalings are given by
\be{hyscale}
b^{0} \rightarrow  b^{0 (b)}  + \frac{1}{m_p} \delta b^{0 }
,\;\;\;\;\;
b^x \rightarrow \frac{1}{m_p} b^x, \;\;\;\;
\xi^\alpha \rightarrow \frac{1}{m_p^{3/2}} \xi^\alpha
\ee
Note that the couplings of the hypermultiplets to the vectors is dictated by
the charges which are scaled as in \eq{scalchy}. This and the above
scaling for the hypermultiplets ensures that $b^x$ and $\delta b^0$ decouples
from the dynamics. $b^{0(b)}$ is the background value which enters as
a coupling constant in the decoupled theory with global supersymmetry.

\section{ Partial breaking of ${\cal N}=2$ in the rigid limit}

In this section we look for vacua with ${\cal N}=1$ supersymmetry in
the rigid limit. In the rigid limit since fluctuations of the
graviton and the hyperino decouple it is sufficient to study the
supersymmetry transformation of the gauginos and look for vacua
which preserve a single supersymmetry.
As we mentioned before we will not keep track of the gravitational background
couplings to the resulting ${\cal N}=1$ theory in the Lagrangian but
will include them in the analysis of the chiral ring relations in
section 5.  In fact including the gravitational background couplings
makes it extremely cumbersome to write a compact form for the
Lagrangian, without them can  summarize the resulting Lagrangians
compactly in ${\cal N}=1$ superspace. 

To analyze the supersymmetry of the vacuum we study the 
 gaugino shifts which are  given by
\bea{gaugshi}
\delta\lambda^{Aa} &=& W^{aAB} \eta_B  + \ldots \\ \nonumber
&=& \left( \epsilon^{AB} k^{a}_\Lambda
\bar{L}^\Lambda + i (\sigma_x)_C ^{\;\;B} \epsilon^{CA} {\cal
P}^x_\Lambda g^{ab*} \bar f^\Lambda_{b*} \right) \eta_B + \ldots
\eea
There are other terms in the supersymmetric variation of the gauginos
which represent the  dots.
But the terms explicitly written down in \eq{gaugshi} are alway
present irrespective of which gravitational backgrounds are turned
on, therefore they dictate the number of supersymmetry
preserved at various vacua. 
Substituting the scalings of all the fields in the above expression we
obtain
\bea{sugauginino}
\delta\lambda^{Ai} &=& \Lambda^2 \epsilon^{AB} \frac{1}{\sqrt{2}} 
f^i_{jk}z^k \bar{z}^j \eta_B \\ \nonumber
&+& i \Lambda^2 \sqrt{2}  \epsilon_{BC}
\frac{1}{ b^0}
 \tau_2^{ia} \left( 
 \xi \sigma_A^{1C} \delta_{a*n*} + 
 (e \delta_{a*n*}  + m \bar{\tau}_{a*n*} )  \sigma_A^{2C}  \right) 
\eta ^B + O (\frac{\Lambda^3}{m_p} ) , \\ \nonumber
\delta\lambda^{An} &=& i \Lambda^2 \sqrt{2} \epsilon_{BC} \
\frac{1}{ b^0} 
 \tau_2^{na} \left( 
 \xi \sigma_A^{1C} \delta_{a*n*} + 
 (e \delta_{a*n*}  + m \bar{\tau}_{a*n*} )  \sigma_A^{2C}  \right) 
\eta ^B + O (\frac{\Lambda^3}{m_p} ) , 
\eea
where in $b^{(0)}$ we have suppressed the superscript to indicate the
background and we have defined
\bea{aptdef}
{\cal F} (z) = i (z^n)^2 - 2i \phi(z), \\ \nonumber
\tau_{ij}(z)  = \tau_1(z)_{ij} + i \tau_2(z)_{ij} = \partial_{i}
\partial_j
{\cal F}(z)
\eea
Now it is easy to see that we  have a vacuum with ${\cal N}=1$
supersymmetry. Consider the following vacuum values
\be{vacval}
z^i =0,  \;\;\;\;\; 
\tau_{ij} = \left( -\frac{e}{m} + i |\frac{\xi}{m}| \right) \delta_{ij} , 
\;\;\;\;
\tau_{in} = 0, \;\;\;\;\;  \tau_{nn} = -\frac{e}{m} + i |\frac{\xi}{m}|
\ee
For these values of the scalar field the shifts on the gauginos reduce to 
\bea{redshg}
\delta \lambda^{Ai} &=& 0, \\ \nonumber
\delta \lambda^{An} &=& i \Lambda^2 \sqrt{2} \epsilon_{BC} 
\frac{1}{ b^0} m ( \sigma^1 + i \sigma^2)_A^C \eta^B
\eea
From this it is clear for the vacuum \eq{vacval} there is one unbroken
supersymmetry as the shift on the gaugino $\lambda^{An}$ has one zero
eigen value. Also note that the scale of the supersymmetry breaking
is controlled by $\Lambda$.
Due to the gauging of the hypermultiplet manifold and the
vector multiplet manifold, there is a
potential for the scalars of the vector multiplet which is given by
\eq{fulscalpot}. 
The leading term in the potential which will contribute in the
$m_p\rightarrow \infty$ limit is given by
\bea{scalpot}
V(z) &=& g^2 \frac{\Lambda^4}{m_p^4} \left(
\frac{1}{4} \tau_{2ij} f^i_{lm} \bar{z}^l z^m f^j_{np} z^n \bar{z}^p
\right. \\ \nonumber
&+& \left. \frac{1}{b^{02}} \tau_2^{nn} \xi^2 
+ \frac{1}{b^{02}} \tau_2^{ab*} ( e \delta_{an} + m \tau_{an} ) ( e
\delta_{b*n*} + m \tau_{b*n*} ) \right) \\ \nonumber
&-& \frac{\Lambda^4}{m_p^4} \frac{1}{2} \left( \xi^2 + e^2 +4m^2
\right)
\eea
The $m_p$ dependence in the scalar potential cancel in the Lagrangian
as the coordinate $x$ is scaled to $x \rightarrow m_p x$.
The first term in the scalar potential arises from the first term of
\eq{fulscalpot}, it corresponds to the scalar potential arising due to the
D-term of a rigid ${\cal N}=1$ theory. The second and third terms of
\eq{scalpot} are due to the coupling of the vectormultiplet manifold to the
gaugings of the hypermultiplet manifold. These terms are obtained by
substituting the apropriate scalings in the third term of the 
potential in \eq{fulscalpot}. The constant contribution to the 
scalar potential in \eq{scalpot} in obtained from the rigid limit of 
the second and the last term of \eq{fulscalpot}.
It is also easy to see that the the vacuum \eq{vacval} is the minimum of the
above scalar potential. The minimum value of the potential is infact
not zero but is given by
\be{minscalpot}
V_{\rm{ min} } = g^2 \frac{\Lambda^2}{2b^{02}} ( 4\xi m - \xi^2 -e^2 -4m^2),
\ee
thus it is clear that the vacuum \eq{vacval} does not preserve
${\cal N}=1$ supersymmetry in the fluctuations of the  gravitational sector
which are decoupled. 
However the gravitational backgrounds do transform into each other
under the full ${\cal N}=2$ supersymmetry transformations. This is
because the constants shifts of the gravitinos are suppressed with
respect to the variation of the backgrounds by $1/m_p$, which is the
same ratio of scales between the background and the fluctuations.
In the subsequent discussion, as we will be interested in the rigid
limit we will, neglect the zero point energy \eq{minscalpot} 
and  will set the background
moduli $b^0 =1$.

\subsection{Structure of the theory about the ${\cal N}=1$ vacuum}

To study the structure of the low energy theory which is obtained  
at the ${\cal N}=1$ vacuum we will
first evaluate the mass matrices for the fermions and show that 
half the fermions are massless and half of them are massive. Thus the
gaugino of the ${\cal N}=2$ vector multiplet splits into the partner
of the ${\cal N}=1$ gauge multiplet and a massive  ${\cal N}=1$ chiral
multiplet in the adjoint representation.
The masses of the fermions are extracted from the term
${\cal M}_{aAbB} \bar{\lambda^{aA}} \lambda^{bB}$ where
\be{defm}
{\cal M}_{aAbB} = \frac{1}{3} \left( \epsilon_{AB} g_{aa*}k^{a*}_\Lambda
f^\Lambda_b + i \sigma_{xA}^C\epsilon_{CB} {\cal P}^x_\Lambda \nabla_a
f^\Lambda_b \right)
\ee
The ${\cal N}=1$ vacuum preserves the $U(N)$ gauge symmetry 
thus the expectation valued $<z^i> =0$, which enables us to drop the
first term in \eq{defm}.
To evaluate the second term in 
the mass matrix we need the Christoffel symbols of the
vector multiplet manifold at the
${\cal N}=1$ vacuum to the leading order in $\Lambda/m_p$. 
Using the scaling in \eq{scaleqan}, 
the non vanishing Christoffel symbols to the leading order in
$\Lambda/m_p$ are
\be{defc}
\Gamma^n_{ab} = -\tau_2^{nn} \partial_{a}\partial_{b}\partial_n
\phi
\ee
Substituting this and using the scaling of \eq{scalef} and  \eq{scalchy} 
we obtain
\be{mass}
{\cal M}_{aAbB} = - \frac{\Lambda^3}{m_p^3}
\frac{1}{6 \sqrt{2} } \left. \epsilon_{AC}
m ( \sigma^1 + i \sigma^2 )^C_B \partial_{a}\partial_b\partial_n \phi
\right|_{z^a=0}
\ee
The $m_p$ scaling of the fermion mass is such that the mass
term  ${\cal M}_{aAbB} \bar\lambda^{aA}\lambda^{bB}$ scales like
$\Lambda/m_p^4$, which is the right scaling so that it survives in the
$m_p\rightarrow\infty$ limit. The fermion mass is proportional to
$\Lambda$, the supersymmetry breaking scale as expected.
Now it is easy to see that this mass matrix has $N^2$ zero 
eigenvalues, the corresponding gauginos being members  
of the ${\cal N}=1$ gauge multiplets.
The $N^2$ non-zero eigenvalues,  give mass proportional to 
$\partial_a\partial_b\partial_n\phi$ to the fermions which are part
of the ${\cal N}=1$ chiral multiplets. Thus from \eq{mass} we see that
the fermions $\lambda^{i1}$ are the gauginos and $\lambda^{i2}$ are
the fermions of the chiral multiplets which  become massive.

The structure of the  ${\cal N}=1$ theory will be dictated by the
superpotential for the chiral multiplets and the gauge function. 
Since the ${\cal N}=1$ theory is obtained by spontaneously breaking
${\cal N}=2$ the gauge function and the superpotential 
are closely related, both of them are derived from a prepotential.
We will see how this happens for the theory we are discussing at the
${\cal N}=1$ vacuum. 
We organize the expansion of the scalar potential in \eq{scalpot} around the
the vacuum in \eq{vacval} using the following form for the expansion
of the gauge function
\bea{expvac}
\tau_{ij} &=& \left( -\frac{e}{m} + i |\frac{\xi}{m}| \right)
\delta_{ij} +  \tau'_{ij}, \\ \nonumber
\tau_{in} &=& \partial_i {\cal W} , \\ \nonumber
\tau_{nn} &=&  -\frac{e}{m} + i|\frac{\xi}{m}| + \partial_n {\cal W}
\eea
This parametrization of $\tau_{in}$ and $\tau_{nn}$ can always be done
because of its definition \eq{aptdef} with  $ {\cal W} = \partial_n {\cal F}$.
Substituting this form into the
expression for the scalar potential in \eq{scalpot} we obtain
\be{onescalpot}
V (z) = g^2 \frac{\Lambda^4}{m_p^4} \left(
\frac{1}{4} \tau_{2ij} f^i_{lm} \bar{z}^l z^m f^{j}_{np} z^n\bar{z}^p
+   m^2 \tau_2^{ab*} \partial_a {\cal W} (\partial_b {\cal W })^* \right) + 
V_{\rm{min} }
\ee
Before we go ahead we will fix the factor $g$ that appears in the
scalar potential. From \cite{Andrianopoli:1997cm}
we see that covariant derivatives of the scalar fields $z$ are given
by
\bea{covdez}
dz^i + g A^\Lambda k^i_\Lambda(z) &=&
dz^i +  g f^\Lambda_k A^k f^i_{\Lambda, j} z^j + \cdots \\ \nonumber
&=& \frac{1}{m_p} ( dz^i +  g \frac{1}{\sqrt{2}} A^k f^i_{kj} z^j +
\cdots )
\eea
Here we have expressed the field $A^\Lambda$ in terms of the gauge
fields $A^i$ and the graviphoton which are represented by the dots.
In writing this equality we have ignored the contribution to the
covariant derivative from the graviphoton as the first term is
sufficient to fix the factor $g$. The second equation in \eq{covdez}
is obtained by substituting the required scalings, now it is clear
that $g=\sqrt{2}$ is a convenient choice so that we obtain the
standard commutator term without extra factors.

We now detail the structure of the ${\cal N}=1$ theory. From the
scalar potential we see that the ${\cal N}=1$ theory has a
superpotential $m {\cal W}$, a nontrivial gauge function $\tau$ and a
nontrivial K\"{a}hler term with a non trivial metric 
for the scalars also given by $\tau$. 
To arrive at the ${\cal N}=1$ theory with a polynomial superpotential 
we choose the following prepotential.
\bea{prepot1}
{\cal F} &=& 
( -\frac{e}{m} + i |\frac{\xi}{m}| ) {\rm Tr}\frac {(z^2)}{2} +
\sum_{k=1}^p \frac{1 }{(k+1)(k+2)}  g_k'  {\rm Tr} ( z^{k+2} )
\\ \nonumber
&=& 
( -\frac{e}{m} + i |\frac{\xi}{m}| ) {\rm Tr} \frac {(z^2)}{2}
+  {\cal F}'
\eea
Here $z$ refers to the $N \times N$ matrix given by
$ z= \frac{z^n}{\sqrt{N}} I + z^i T_i$. 
We have chosen the prepotential so that it
satisfies the condition that, at $z^a=0$, one has the
required vacuum values as in \eq{vacval}. 
Thus from comparison of the expansion in \eq{expvac} and the scalar
potential in \eq{onescalpot}
the superpotential is given by
\bea{defw}
{\cal W} &=&  m \partial_n {\cal F}' = 
\frac{m}{\sqrt{N}} 
\sum_{k=1}^p \frac{1}{k+1} g_k {\rm Tr} (z^{k+1}), \\ \nonumber
&=& \tilde{m} 
\sum_{k=1}^p \frac{1}{k+1} g_k {\rm Tr} (z^{k+1}). 
\eea
In the second equality we have absorbed a factor of $\sqrt{N}$ by the 
redefinition  $m = \tilde{m} \sqrt{N}$ for convenience. 
The gauge kinetic function is given by
\bea{gaugfn}
\tau_{ij} &=& (-\frac{e}{m} + i |\frac{\xi}{m}| )\delta_{ij} +  
\partial_{ij} {\cal F}', \\ \nonumber
\tau_{in} &=& \partial_{in} {\cal F}', \\ \nonumber
\tau_{nn} &=& -\frac{e}{m} + i |\frac{\xi}{m}| + \partial_{nn}{\cal F}'
\eea
Again from the expansion of the scalar potential we see that  
the K\"{a}hler metric for the scalars is given by $\tau_{2 ij}/2$

Finally, before writing the resulting action in superspace we will 
restore the canonical dimension of the scalar 
field $z$ by scaling $z\rightarrow z/\Lambda$ and as a result given
dimensions to the couplings in the superpotential. 
To extract the
canonical dimension of the superpotential, one scales
${\cal W} \rightarrow {\cal W}/\Lambda^3$. 
Doing this would require the scaling of the
couplings $g_k \rightarrow    \Lambda^{k-2}  g_k$, this ensures that
that the scaled superpotential is given by \eq{defw} with the
understanding that the field $z$ and the couplings have the appropriate
dimensions. 
However, the gauge kinetic function contains one more
derivative with respect to  $z$, and therefore 
performing the same scaling of  $z$ and the couplings  in the gauge
kinetic function  we obtain the following scaling forms
\bea{gaugfns}
\tau_{ij} &=& (-\frac{e}{m} + i |\frac{\xi}{m}| )\delta_{ij} +  
\frac{1}{\Lambda^2} \partial_{ij} {\cal F}', \\ \nonumber
\tau_{in} &=& \frac{1}{\Lambda^2} \partial_{in} {\cal F}', \\ \nonumber
\tau_{nn} &=& -\frac{e}{m} + i |\frac{\xi}{m}| + \frac{1}{\Lambda^2} 
\partial_{nn}{\cal F}'
\eea
Note that if one  is interested in the dynamics of the theory at
scales much smaller that the 
supersymmetry breaking scale $\Lambda$ we obtain a ${\cal N}=1$ theory
with a trivial gauge kinetic function and a trivial K\"{a}hler metric
for the scalars leading to theory discussed in
\cite{Dijkgraaf:2002dh}.  We summarize the ${\cal N}=1$ theory
obtained from partial breaking of ${\cal N}=2$ in superspace. 
\bea{ftheory}
S &=& -\frac{i}{4} \int d^2\theta \tau_{ab} (\Phi) W^a W^b + 
 \int d^2 \theta {\cal W} (\Phi)  + S_{ \rm{ Kahler} }, \\ \nonumber
S_{ \rm{ Kahler}} &=& \int d^4 \theta K (\Phi, \Phi^\dagger), \\
\nonumber
K(z, \bar{z}) &=&   \frac{i}{4} \left( z^a \partial_{a*} \bar {\cal F} -
\bar{z}^a  \partial_a {\cal F} \right)
\eea
Here the K\"{a}hler potential is of dimension 2 and we have used the
scaled $z$ and the scaled couplings. 
This is the non-Abelian generalization of the ${\cal N}=1$ model obtained  by 
\cite{Antoniadis:1996vb} for partial breaking in rigid ${\cal N}=2$ theories. 

\section{Partial breaking with hypermultiplets in the rigid limit}

As we have seen in the previous section we need at least one  
hypermultiplet in the hidden
sector in order to break ${\cal N}=2$ supergravity. 
The low energy theory in
the rigid limit was ${\cal} N=1$ with an adjoint chiral multiplet. 
In this section we would like to generalize the discussions of the
previous section to obtain chiral multiplets charged in the
fundamental representation of the gauge group in the rigid limit 
of spontaneously broken
${\cal N}=2$ supergravity. We will discuss the  case of obtaining chiral
multiplets charged in the fundamental of $U(N)$ in detail and then
outline the case for other representations.

We discuss the field content and the gaugings necessary in detail.
We start with a $n+1$ dimensional special K\"{a}hler manifold for
the vector multiplet. The holomorphic sections on this manifold is
given by
\bea{vecsec1}
X^{0}(z)  = \frac{1}{\sqrt{2}}, \;\;&\;&\;  
F_{0}(z) = -\frac{i}{\sqrt{2} }
\left( 2 f(z) - z^a \frac{\partial f( z) }{\partial z^a} \right) , \\
\nonumber
X^i(z) =   \frac{z^i}{\sqrt{2}} 
\;\;&\;&\; F_i (z) = 
- \frac{i}{\sqrt{2}} \frac{\partial f(z) }{\partial z^i}, 
 \\ \nonumber
X^{n+1}(z) = \frac{i}{\sqrt{2} } \frac{\partial f(z) }{\partial z^{n+1}}, 
\;\;&\;&\;
F_{n+1}(z) = \frac{z^{n+1} }{\sqrt{2}}. 
\eea
Note that, other than for the inclusion of an additional coordinate, 
this  holomorphic sections is similar to \eq{vecsec}. Here $i, j$ take
values from $1, \ldots n$,  and $a$ takes values from $1$ to
$(n+1)=m$. 
The $i, j$ indices parametrize the $U(N)$
directions, with $n=N^2$, while the $m$ th direction corresponds to
an additional $U(1)$.  Thus we start with the gauge group $U(N)\times
U(1)$. We need this additional $U(1)$ as it is not possible to obtain
hypers with respect to the single $U(1)$ for ${\cal N}=1$ obtained by 
partially breaking ${\cal N}=2$ \cite{Partouche:1997yp}. Thus we need
an additional $U(1)$ to provide $U(1)$ charges for the hypers. 
In the spontaneously broken theory the $U(1)$ responsible 
for the gauging which partially breaks the supersymmetry 
will be decoupled. 
The action of the group $U(N)$ on the coordinates $z^i$ is the given
by
$U(z^i T_i ) U^{\dagger}$, here $T_i$ stand for the the $N\times N$
Hermitian matrices which generate the full group $U(N)$. 
The section \eq{vecsec1} satisfy the special K\"{a}hler condition given
in \eq{skcond}. The K\"{a}hler potential and the Killing vectors are
given by similar equations as \eq{kahpot} and \eq{kill}, except that
the structure group is $U(N)$ and the coordinates $i, j, k$ run from
$1$ to $N$. 

To obtain matter in the fundamental representation of $U(N)$ it is
convenient to take the hypermultiplet manifold to be  the following
homogeneous, symmetric, quaternionic  manifold
\be{quat}
{\cal M} = \frac{ U(2, N+1 )}{U(2)\times U(N+1)}
\ee
This manifold has dimension $4(N+1)$. Indeed we 
need one additional hypermultiplet to play  the role
of  $SO(4,1)/SO(4)$  of the previous section. 
this is the minimal requirement for spontaneous breaking of ${\cal N}=2$
supergravity, however here the hypermultiplet  is non-trivially embedded
in ${\cal M}$. 
We parametrize this coset manifold using the following
$(3+ N)\times (3+N)$ matrix
\be{parcos}
{\cal L} (q) =  \left(
\begin{array}{c c}
\sqrt{ 1 + q q^{\dagger} } & q \\ 
q^\dagger & \sqrt{ 1 + q^\dagger q} 
\end{array}
\right) = 
\left(
\begin{array}{ c c}
r_1  & q  \\ 
q^\dagger & r_2 
\end{array}
\right)
\ee
here $q$ is a $2\times (N+1)$ matrix with complex entries. We label the
entries as $q^{Au}$, $A$ running over $1, 2$ and $u$ running over $1, \ldots
m$. Note that this parametrization ensures that ${\cal L}$ satisfies
\bea{condl}
{\cal L} \Omega {\cal L}^{\dagger} = \Omega , \\ \nonumber
\Omega =  \left(
\begin{array}{ c c }
-1_{2\times 2}  & 0 \\
0 & 1_{(N+1) \times (N+1)} 
\end{array}
\right)
\eea
From this parameterization, the coset vielbein $E$, the $U(2)$
connection $\theta$ and the $U(N+1)$ connection $\Delta$ are given by
\bea{vcc}
E &=& r_1dq - q dr_2 = dq - \frac{1}{2} q dq^\dagger dq + O(q^4), \\
\nonumber
\theta &=& r_1 dr_1 - q dq^\dagger =  \frac{1}{2} ( dqq^\dagger - q
dq^\dagger) + O(q^3), \\ \nonumber
\Delta &=& r_2 dr_2 - q^\dagger dq = \frac{1}{2} ( dq^\dagger q -
q^\dagger dq)  + O(q^3)
\eea
The metric on the quaternionic manifold ${\cal M}$ is given by
\be{metq}
ds^2 = E ^\dagger \otimes E = dq^\dagger dq  -\frac{1}{2} \left(
q^\dagger dq q^\dagger dq + dq^\dagger q dq^\dagger q \right)
+ O(q^4)
\ee
We have written down the expansion in powers of $q$  as it will  be
useful for us later, in taking the $m_p\rightarrow \infty$ limit.
The HyperK\"{a}hler 2-form $K^x$ and the $SU(2)$ triplet 1-forms are
given by
\be{hypst}
K^x = \frac{1}{2} {\rm Tr} ( E^\dagger \wedge \sigma^x E ),
\;\;\;\;\;
w^x = -\frac{1}{2} {\rm Tr} ( \theta \sigma^x)
\ee
where $\sigma^x$ refers to  Pauli spin matrices.
We denote the  vielbein 1-form as ${\cal U}^{Aai}$, where we have
split the $Sp(2N+2)$ indices as $ai$ with $a= 1,2$ and $i= 1, \ldots
N+1$. The components of the vielbein 1-form are defined by 
\bea{defviel}
{\cal U}^{Aai} &=& {\cal U}^{Aai}_{Bu} dq^{Bu} + 
{\cal U}^{Aai}_{B\bar{u}} d \bar{q}^{B\bar{u}} , \\ \nonumber
&=& \delta^{a1} \epsilon^{AB} \delta^{ij} E^{Bj} 
+ \delta^{a2} \delta^{AB} \delta^{ij} E^{\dagger jB}
\eea

We now write down the Killing vectors corresponding to the isometries
of the manifold ${\cal M}$. For every element $g\in u(2, N+1)$, the 
Lie algebra of $U(2, N+1)$, the
following action is an isometry 
\bea{kilac}
\delta{\cal L} \equiv  (k_g^{Au} \frac{\partial}{\partial q^{Au}} 
+ k_g^{A\bar{u} }\frac{\partial}{\partial q^{A\bar{u} } }){\cal L}
= g{\cal L} - {\cal L} w_g
\eea
where $g \in u(2, N+1)$. $w_g \in u(2)\oplus u(m)$ is
the right-compensator, which depends on $g$ and is needed
to keep $\delta {\cal L}$ in the form  of (\ref{parcos}).
$g$ and $w_g$ are parametrized by
\be{gandwg}
g= \left( 
\begin{array} {c c}
a & b \\ 
b^\dagger & c
\end{array}
\right) \in u(2, N+1) , \;\;\;\;\;
w_g = \left(
\begin{array}{c c}
w_1 & 0 \\
0 & w_2 
\end{array}
\right) \in u(2) \otimes u(N+1)
\ee
with $a$ and $b$ anti-hermitian matrices. 
Expanding \eq{kilac} out we get
\bea{constoq}
\delta q = aq + br_2 - qc - q \hat{w}_2  &=& aq + r_1b -qc + \hat{w}_1
q, \\ \nonumber
w_1 = q+ \hat{w}_1, \;\;\;&\;&\;\;\;  w_2 = c + \hat{w}_2
\eea
To leading order  the solutions for $\hat{w}$ 's in terms of $g$ is
is given by
\be{solw}
\hat{w}_1 = \frac{1}{2}( bq^\dagger - qb^\dagger ) + O(q^3) , \;\;\;\;
\hat{w}_2 = \frac{1}{2} (b^\dagger q - q^\dagger b) + O(q^3)
\ee
Using all this killing vectors to the leading order in $q$ for a
given $g$ is given by
\bea{hykill}
\vec{k}_{a} &=& (aq)^{Au} \frac{\partial}{\partial q^{Au} } +
(q^\dagger a^\dagger)^{\bar{u} A} 
\frac{\partial}{\partial q^{\dagger A \bar{u}}}, \\ \nonumber
\vec{k}_{a} &=& -(qc)^{Au} \frac{\partial}{\partial q^{Au} } 
-(c^\dagger q^\dagger)^{\bar{u} A} 
\frac{\partial}{\partial q^{\dagger A \bar{u}}}, \\ \nonumber
\vec{k}_{a} &=& (b)^{Au} \frac{\partial}{\partial q^{Au} } +
( b^\dagger)^{\bar{u} A} 
\frac{\partial}{\partial q^{\dagger A \bar{u}}}, 
\eea
where we have used the subscripts to denote the entry of $g$.
Therefore
Killing vector for the group element $g$ is given by  $\vec{k}_g = \vec{k}_a +
\vec{k}_c + \vec{k}_b$. 
From the Killing vectors  is easy to compute the moment map, which is
defined by the equation
\be{defmom}
{ \bf i}_{\vec{k}_g} K^x = -\nabla {\cal P}^x
\ee
From the fact that the curvature of the $SU(2)$ connection is
proportional to the HyperK\"{a}hler structures, we can solve the
above equation by the ans\"{a}tz
\be{solmom}
{\cal P}^x_g  = \frac{1}{2} {\rm Tr} \left[ \left( 
\begin{array}{ c l }
\sigma^x & \;\;\; 0 \\
0 & \;\;\; 0_{(N+1)\times (N+1)} 
\end{array} \right) 
L^{-1} g L  \right]
\ee
For later convenience we write down the leading contribution to the 
moment maps
from $a$, $b$ and $c$ of the isometry $g$. 
\bea{expsolmom}
{\cal P}^x_{a}  &=& \frac{1}{2} {\rm Tr} 
\left( \sigma^x a + \frac{\sigma^x }{2} (   qq^\dagger a + a qq^\dagger)
\right), \\ \nonumber
{\cal P}^x_{b} &=& - \frac{1}{2} {\rm Tr}
\left( \sigma^x qb^\dagger - \sigma^x bq^\dagger \right) , \\
\nonumber
{\cal P}^x_{c} &=& - \frac{1}{2} {\rm Tr} 
\left( \sigma^x qcq^\dagger \right)
\eea

We now give the two commuting matrices,  which belong to the Lie
algebra $u(2, N+1)$, which are the generators of $R^2$. We will gauge
along this isometries to break supersymmetry. These Killing
directions function as the two translation isometries of the minimal
hypermultiplet manifold $SO(4,1)/SO(4)$ that we used in the previous
section to break supersymmetry. 
\bea{commat}
g_0 &=&  i \left(
\begin{array}{ c c c c c }
\frac{\xi}{2}  & i\frac{e}{2} - \frac{\xi}{2}  
& \frac{1}{2\sqrt{2}} ( e + 2i \xi )  
& 0 &\cdots \\ 
-i \frac{e}{2} -  \frac{\xi}{2}  & 
\frac{\xi}{2}  &   \frac{1}{2\sqrt{2}} ( e - 2i \xi )  
& 0 &\cdots \\ 
 \frac{1}{2\sqrt{2}} ( -e + 2i \xi )  
& \frac{1}{2\sqrt{2}} ( -e - 2i \xi )   
& -\xi & 0 &\cdots \\
0 & 0 & 0 & \;\;  & \;\; \\
\vdots &\; &\;& \;\; &  M I_{N\times N}  
\end{array}
\right) \\ \nonumber
g_{n+1} &=& i \left(
\begin{array}{ c c c c c }
0 & im & \frac{m}{\sqrt{2}}   & 0 &\cdots \\ 
-im  & 0 & \frac{m}{\sqrt{2}}  & 0 &\cdots \\ 
-\frac{m}{\sqrt{2}}  & -\frac{m}{\sqrt{2}}  & 0 & 0 &\cdots \\
0 & 0 & 0 & 0 &\cdots \\
\vdots &\; &\;& \vdots & \cdots
\end{array}
\right) 
\eea
Here the parameters $\xi, e, m $ are real, and it is easy to see that these
isometries commute. Both of them contain a non zero $b$ component,
thus they generate translations also, 
these isometries are used to break supersymmetry. The remaining
isometries, which ensure that the hypermultiplets surviving the
supersymmetry breaking in the rigid limit are gauged with respect to
the $U(N)$ gauge fields, are given by
\be{suvgaug}
g_i = \left(
\begin{array}{ c c c c c }
0 & 0 & 0 &  0 &\cdots \\
0 & 0 & 0 &  0 &\cdots \\
0 & 0 & 0 &  0 &\cdots \\
0 & 0& 0& 0 & \; \; i T^i _{N\times N}
\end{array}
\right)
\ee
here  $T^i$ are  $N\times N$ Hermitian generators of $U(N)$. 

We provide the various scalings necessary in order to 
obtain ${\cal N}=1$ supersymmetry in the rigid limit. 
The scalings of the vector multiplet manifold are given by
\bea{newvecsc}
f(z) = \frac{1}{2} + \frac{\Lambda}{m_p} z^{m} +
\frac{\Lambda^2}{m_p^2} \phi(z, z^m), \\ \nonumber
X^i \rightarrow \frac{\Lambda}{m_p}, \;\;\; F_i \rightarrow
\frac{m_p}{\Lambda} F_i
\eea
The gaugings in \eq{commat} are scaled as follows
\be{scalcom}
\xi \rightarrow \frac{\Lambda^2}{m_p^2}\xi, \;\;\; e \rightarrow
\frac{\Lambda^2}{m_p^2} e, \;\;\; m \rightarrow \frac{\Lambda}{m_p} m,
\;\;\;
M \rightarrow \frac{\Lambda}{m_p} M
\ee
The gravity multiplet and the vector multiplet scale as in
\eq{gfscale} and \eq{vscale} respectively
The hypermultiplet fields scale as 
\be{scalnewhy}
q^u \rightarrow \frac{1}{m_p} q^u, \;\;\; \zeta^\alpha \rightarrow
\frac{1}{m_p^{3/2}} \zeta^\alpha
\ee
where we denote by $\zeta$ the hyperinos, partners of the $q$'s.
The supersymmetry variation with these scalings  are given by
\bea{susgaughy}
\delta\lambda^{Ai} &=& \Lambda^2 \epsilon^{AB} \frac{1}{\sqrt{2}} f^{i}_{jk}
z^k \bar{z}^j \eta_B \\ \nonumber
&-&i {\sqrt{2}} 
\Lambda^2  \epsilon_{CA} \tau_2^{ia} (\sigma_x)_C^{\;\;B}
\left( {\cal P}^x_0 \delta_{a*n*} + 2 {\cal P}^x_m \bar{\tau}_{a*n*}
\right) \eta^B + O(\frac{\Lambda^3}{m_p}), \\ \nonumber 
\delta\lambda^{An} &=& 
- i {\sqrt{2} }
\Lambda^2  \epsilon_{CA} \tau_2^{na} (\sigma_x)_C^{\;\;B}
\left( {\cal P}^x_0 \delta_{a*n*} + 2 {\cal P}^x_m \bar{\tau}_{a*n*}
\right) \eta^B + O(\frac{\Lambda^3}{m_p})
\eea
It is clear that we can still choose the vacuum values for the vector
multiplet moduli to be given similar to \eq{vacval}.
\be{susvac}
z^i =0, \;\;\; 
\tau_{ij} = ( -\frac{e}{m} + i |\frac{\xi}{m}| ) \delta_{ij}, \;\;\;\;
\tau_{im} = 0, \;\;\; \tau_{mm} = - \frac{e}{m} + i |\frac{\xi}{m}|
\ee
Substituting these values in the first equation of \eq{susgaughy} we
see that variation $\delta \lambda ^{Ai} =0$. To see the reduction
of  supersymmetry down to ${\cal N} =1$,
 we consider the last second equation of \eq{susgaughy}, after
substituting the values of the moment maps, we see that the only 
contribution comes from ${\cal P}^x_a$, the others being 
suppressed by
$1/m_p$. The variation is given by
\be{vacvarg}
\delta\lambda^{An}= 
{\sqrt{2} } \Lambda^2 \epsilon_{AC}
m ( \sigma^1 + i \sigma^2)_C^{\;\;B} \eta^B
\ee
Therefore we see now that this vacuum has one unbroken
supersymmetry.

We expand the theory around this vacuum and write down the structure
of the resulting ${\cal N}=1$ theory.
The leading terms in the scalar potential,
which survive in the rigid limit is given by.
\bea{fullscalpot}
V &=& g^2( V_1 + V_2 + V_3 + V_4), \\ \nonumber
V_1 &=& \frac{\Lambda^4}{m_p^4} \left(
\frac{1}{4} \tau_{2ij} f^i_{lm} \bar{z}^l \bar{z}^m f^j_{np}
z^n\bar{z}^p  \right.  \\ \nonumber
&+&\left. \tau_2^{nn} (\frac{\xi}{2})^2 
+ \tau_2^{ab} 
( \frac{e}{2} \delta_{am} + \frac{m}{2} \tau_{am})
( \frac{e}{2} \delta_{bm} + \frac{m}{2} \bar{\tau}_{bm}) \right) \\
\nonumber
V_2 = &=& \frac{1}{m_p^4} \frac{1}{4} \tau_2^{ij} {\rm Tr} ( \sigma^x
qT_i q^\dagger ) {\rm Tr} ( \sigma^x q T_j q^\dagger )  \\ \nonumber
&+& \frac{\Lambda^2}{m_p^4} \frac{1}{2} \tau_2^{mi} \left( \xi {\rm
Tr}( \sigma^1 qT_iq^\dagger)  + e {\rm  Tr}( \sigma^2 qT_iq^\dagger)
\right) \\ \nonumber
&+& \frac{\Lambda^2}{m_p^4} \frac{m}{4} \tau_2^{ai} \left( \tau_{ma} +
\bar{\tau}_{ma} \right) {\rm Tr} ( \sigma^2 q T_i q^\dagger) \\
\nonumber
V_3 &=&  \frac{\Lambda^2}{m_p^4} \left(  2 {\rm Tr} (q M^2 q^\dagger) 
+ 2 {\rm Tr}( q M T_i q^\dagger) ( z^i + \bar{z^i}) \right) \\
\nonumber
V_4 &=& - \frac{\Lambda^4}{m_p^4} (e^2 - 5\xi^2 + 4m^2)
\eea
Note that the first hypermultiplet,  
$q^{A1}$, is massless, and it does not
contribute to the scalar potential.  Thus it plays the same role as
the modulus $b_0$ in the discussion of the previous
section. We now expand this potential around the minima \eq{susvac},
adopting the  same form for the expansion of the gauge function as in
\eq{expvac}
\be{expmin}
\tau_{ij} = ( -\frac{e}{m} + i |\frac{\xi}{m}| ) 
\delta_{ij} + \tau'_{ij}, \;\;\;\;
\tau_{im} = \partial_i {\cal W},  \;\;\; \tau_{mm} = -\frac{e}{m} + i
|\frac{\xi}{m}| + \partial_m {\cal W},  
\ee
Substituting these expansions in the scalar potential
\eq{fullscalpot} we obtain
\bea{potarmin}
V _1 &=& ( \frac{\Lambda}{m_p})^4 \left(
\frac{1}{4} \tau_{2ij} f^i_{lm} \bar{z}^l z^m f^j_{np} z^n \bar{z}^p
+ 2m\xi + (\frac{m}{2})^2 \tau_2^{ab*} \partial_a {\cal W } (\partial_b
{\cal W})^* \right), \\ \nonumber
V_2 &=& \frac{1}{m_p^4} \left( \frac{1}{4} \tau_2^{ij}  
( q^1T_i q^2 - q^{2 \dagger} T_{j} q^{1\dagger})^2 +
(q^1 T_{i} q^2 ) (q^{2\dagger} T_j q^{1\dagger} )  \right. \\ \nonumber
&+& \left. i \frac{m\Lambda^2}{2} \tau_2^{ji} \partial_j 
{\cal W} (q^{2\dagger} T_i
q^{1\dagger} )
- i \frac{m\Lambda^2}{2} \tau_2^{ji} (\partial_j {\cal W})^*  ( q^1 T_i q^2)
  \right), 
\\ \nonumber
V_3 &=& \frac{\Lambda^2}{m_p^4} \left( 2M^2 q^{1i*} q^{1 i} + 2M^2 
q^{2i*} q^{2i} + 
2 ( q^{1} M T_i q^{1\dagger}  +  q^{2\dagger} M T_i q^{2} )
(z^i + \bar{z}^i )  \right.  \\ \nonumber
&+& \left.
2 q^{1} \{T_i, T_j\} q^{1\dagger} z^i \bar{z}^j + 
2 q^{2\dagger} \{T_i, T_j\} q^{2} z^i \bar{z}^j \right)
\eea
Here we have split the ${\cal N}=2$ 
hypermultiplets into two ${\cal N}=1$ 
chiral multiplets $(q^1, q^{2\dagger})$, thus if $q^1$  transforms in
the fundamental representation of $U(N)$ $q^2$ transforms in the
anti-fundamental representation of $U(N)$. It is clear from the
structure of the terms in the scalar potential in \eq{potarmin}, this
transformation property of the hypermultiplets is obeyed.

Using the fact that we have ${\cal N}=1$ symmetry and the information
from the full scalar potential, we can write down the action for the
spontaneously broken theory in a manifestly ${\cal N}=1$ invariant
way, i.e. in superspace. For a theory with a polynomial superpotential
for $z$
we choose the following prepotential
\be{hyprepot}
{\cal F} = ( -\frac{e}{m} + i |\frac{\xi}{m}| ) {\rm Tr}\frac {( z^2)}{2}
+ z^m \sum_{k=1}^p \frac{1}{k+1} g_k' {\rm Tr} ( z^k)
\ee
The prepotential in \eq{hyprepot} satisfies the condition that at
$z^a$, one has the required vacuum values as in \eq{susvac}.
We can also consider adding a term independent of $z^m$ in the above
superpotential, but we will restrict ourselves the the form in
\eq{hyprepot}. The superpotential is given by 
${\cal W}= \frac{m}{2} \partial_m {\cal F}$. Now in addition to performing the  similar scalings as in section
3, to restore the
canonical dimensions of the scalars $z$ and the superpotential we have
to redefine $M \rightarrow M/\Lambda$. 
As in the previous section it will lead to a theory with a superpotential
independent of $\Lambda$ but with a non-trivial $\Lambda$ dependent
gauge kinetic term and a K\"{a}hler term for the scalars. 
Using the scalar potential in \eq{potarmin} we can summarize the
resulting theory in ${\cal N}=1$ superspace language if we set the 
background ${\cal N}=2$ supergravity fields to zero for convenience. 
The ${\cal N}=2$ vector multiplet breaks up into a ${\cal N}=1$ gauge
multiplet $W$ and a massive ${\cal N}=1$ chiral multiplet $\Phi$ in the
adjoint representation of $U(N)$.  The ${\cal N}=2$ hypermultiplets
break up into two ${\cal N}=1$ chiral multiplets $Q_1$ and $Q_2$ in
the fundamental and anti-fundamental representation of $U(N)$. 
The ${\cal N}=1$ superspace effective action, obtained from spontaneously
broken ${\cal N}=2$ action, is given by
\bea{susact}
S &=& S_{{\rm Potential} } + S_{\rm{ Kahler} } + S_{\rm{Gauge}}
+ S_{{\rm Hyper}},
\\ \nonumber
S_{\rm{Potential}}  &=& \int d^2 \theta \left(
{\rm Tr} ( Q_1 \Phi Q_2  + \frac{m}{2} {\cal W}(\Phi) + {\rm Tr}(
Q_1 M Q_2)  \right), \\ \nonumber
S_{{\rm Gauge}} &=& 
\int d^2 \theta \tau _{ab} (\Phi, \Phi, ^m)  W^aW^b , \\
\nonumber
S_{{\rm Kahler} } &=& 
\int d^4 \theta  {\cal K}(\Phi, \Phi_m ; \Phi^\dagger, \Phi^\dagger_m )
\\ \nonumber
S_{{\rm Hyper}} &=& \int d^4 \theta (Q_1^\dagger Q_1 + Q_2^\dagger
Q_2)
\eea
Note that $a, b$ run over $1, \ldots N+1$, which include the $U(1)$
gauge multiplet under which the chiral multiplets 
$Q_1$ and $Q_2$, in the $\bar {\bf N}$, ${\bf N}$ of
$U(N)$ respectively,  are  not charged.
We have also explicitly indicated the dependence of the gauge function
$\tau$ and  the K\"{a}hler kinetic term on the 
chiral multiplet corresponding to $z^m$, the K\"{a}hler metric is
given by $\tau_{ab}$. The kinetic term for the hypermultiplets is the
standard one as in the $m_p\rightarrow \infty$ limit the metric on the
quaternionic space is flat \eq{metq}.
It is clear that the superpotential does not depend on $\Phi_m$ for
the choice of the prepotential given in \eq{hyprepot}.
Note that in the superpotential chiral multiplets coming from the
hypers couple to those from the vectors only linearly. This is
a consequence of the fact that this theory is obtained from ${\cal
  N}=2$ theory.

Although in this section we have focused on obtaining chiral
multiplets in the fundamental and anti-fundamental representation of 
$U(N)$ gauge group, the generalization to arbitrary (complex)
representation is straightforward \footnote{The resulting models are
always vector-like, i.e. if a chiral multiplet transforms in a complex
representation then there is another chiral multiplet in the complex
conjugate representation.}. Indeed, for an $R$-dimensional
representation of $U(N)$, we can start with the quaternionic manifold
$U(2, 1+R)/U(2) \times U(1+R)$ for the hypermultiplets. The gauging
now is done exactly as above with the representation matrix
$T^i_{N\times N}$ in the eq.(\ref{suvgaug}) for $g_i$ replaced by
representation matrices $T^i_{R\times R}$. Since the representation
is unitary it is clear that $T^i_{R\times R}$ are in the Lie algebra
of $U(R)$. Furthermore, if the representation $R$ is reducible, then we
can include different masses for different irreducible factors
in the equation (\ref{commat}) for $g_0$. 
The rest of the analysis goes exactly as above leading
upto the ${\cal N}=1$ action (\ref{susact}), where now $\Phi$
appearing in the term $Q_1 \Phi Q_2$ is in the $R$-dimensional
representation.

Generalization to other gauge groups is also straightforward. For a
gauge group $G$ one starts with a symplectic section $(X^{\Lambda},
F_{\Lambda})$ exactly as given in (\ref{vecsec1}) with $n= {\rm dim}(G)$. 
The K\"{a}hler potential and the Killing vectors are the same as in eqs.
(\ref{kahpot}) and (\ref{kill}), except that the structure group is
$G$ and the indices $i,j,k$ run from $1$ to ${\rm dim}(G)$. One can
also get matter in arbitrary unitary representation $R$ of group
$G$ by starting with the quaternionic manifold given in the previous
paragraph and replacing $T^i$'s by the corresponding representation of
$G$. In particular we can get quiver theories by choosing $G$ to be the
product of $\prod_{a=1}^k U(N_a)$ modulo the center of mass $U(1)$
(the role of the center of mass $U(1)$ is played here by one of the $U(1)$'s
which gauges a translational symmetry under which the remaining hypers
are neutral). We can get the bi-fundamental chiral fields by gauging a
suitable quaternionic manifold. For example, for $A_k$ quiver theory,
we can choose the quaternionic manifold $U(2,n+1)/U(2)\times U(n+1)$
with $n= \sum_{a=1}^{k-1} N_a N_{a+1}$ and gauge in an obvious way
by means of suitable representation matrices $T^i$. In fact, the
example discussed in detail in this section, is an example of $A_2$ 
quiver theory with the gauge group $U(N)\times U(1)$, where the center
of mass $U(1)$ is the extra $U(1)$ that gauges one of the translational
symmetries of the hypers.

\section{Chiral ring relations from ${\cal N} =2$} 

In this section we derive the 
chiral ring relations from the solutions of the Bianchi identities of
${\cal N}=2$ supergravity. 
We look for the appropriate solutions of the Bianchi identities 
from \cite{Andrianopoli:1997cm} and expand them around the vacuum
which partially breaks supersymmetry, retaining the relevant ${\cal
N}=2$ gravity backgrounds. 
Our discussion will be focused on the case of ${\cal N}=2$ 
gravity coupled to $U(N)$ vector multiplets along with the minimal
hypermultiplet responsible for partial breaking of supersymmetry, 
this was the case discussed in section section 3. We will briefly
indicate the derivation of the chiral ring relations for matter in the
fundamental representation. 

Before we go into the details, 
we will first indicate the strategy followed in deriving the chiral
the ring relations.  Consider  a ${\cal N}=2$ chiral superfield 
$V^{(M_1, M_2, \ldots)}$, here the symbols $(M_1, M_2, \ldots)$ refer
to the either internal symmetries or to
bosonic and  spinorial indices in superspace. 
We obtain relations in the chiral ring  
of ${\cal N}=1$ supersymmetry by considering the following 
$\bar{D}$ exact quantity
\be{dext}
\bar{D}^{1\dot{\alpha} }( D_{\alpha\dot{\alpha}} V^{(M_1, M_2 ,
\ldots)}) = [ \bar{D}^{1\dot{\alpha}} , D_{\alpha\dot{\alpha}} ]
V^{(M_1, M_2, \ldots)} 
\ee
Here the 1 in superscript of the covariant derivative refers to the
unbroken ${\cal N}=1$ supersymmetry.  To write down the above
equality we have used the fact that the superfield $V$ is  chiral,
thus the commutator of the covariant derivatives
in \eq{dext} can be set to zero in the ${\cal N}=1$ chiral ring. 
From the definition of covariant derivatives in superspace we have
the following 
\bea{comcor}
( {\cal D}_N {\cal D}_O &-& (-1)^{no} {\cal D}_O {\cal D}_N )
V^{(M_1M_2\ldots)} \\ \nonumber
&=& - R^{M_1}_{NOP}V^{(PM_2\ldots)} - 
R^{M_2}_{NOP}V^{(M_1D\ldots)} -  \cdots
- T^P_{NO} {\cal D}_P V^{(M_1M_2\cdots)}
\eea
Thus the above combination of torsions, curvatures acting on a chiral
superfield vanishes in the chiral ring, providing the relation we
are looking for. To make our job simpler we will actually look at the
commutator of covariant derivatives acting on the bottom component of
a ${\cal N}=1$ superfield and then promote the identity we obtain to
a ${\cal N}=1$ superfield relation. 
To evaluate the torsion and the curvatures involved in identities like
\eq{comcor} we appeal to the
solutions of the Bianchi identities found in \cite{Andrianopoli:1997cm}.
Consider the  following Bianchi identity 
\be{bia}
\nabla ^2  V^M = V^N R_{N}^M,
\ee
here $V^M$ is a superfield valued form in ${\cal N}=2$ superspace,
$\nabla$  is the covariant derivative. Given a set of vielbein's 
in ${\cal N}=2$ superspace we can expand the covariant derivative as 
$\nabla = E^M {\cal D}_M$, where $E^M$ are the supervielbein's in 
${\cal} N=2$ superspace. For example the $\theta =0$ components of the
supervielbein in ${\cal N}=2$ superspace can be chosen to be 
$E^M = (V^a_\mu dx^\mu , \psi_\mu^A dx^\mu, \psi_{\mu A} dx^\mu)$.
\footnote{We have followed the notation of \cite{Andrianopoli:1997cm}
to indicate the chiralities of the spinors. For the gravitinos, 
$\psi^A$ is antichiral and $\psi_A$ is chiral.}
In \eq{bia} the curvature $R_N^M$ is a two form valued in superspace.
Expanding the left hand side in flat indices we get
\bea{flabi}
V^N R_N^M &=& {\nabla }( E^N {\cal D}_N V^M ),  \\ \nonumber
&=& E^N {\nabla } ( {\cal D}_N V^A ) + {\nabla }( E^N) {\cal D}_N V^M,
\\ \nonumber
&=& E^NE^O {\cal D}_O {\cal D}_N V^M + T^N{\cal D}_N V^O
\eea
Note that in the last equality in the above equation one has the
torsion piece and the commutator of the covariant derivatives.
We can obtain the commutator relevant to \eq{dext} by looking for
specific components in the above equation, namely we look for
components with one spatial index, the $\alpha\dot{\alpha}$ component
and one antichiral index in the ${\cal N}=1$ superspace.
The equations (A.6) to (A.18) along with (A.26), (A.27) and (7.58) 
of \cite{Andrianopoli:1997cm}  provide the
information of the curvatures in \eq{flabi}, while the torsions can
be read from equations (A.23) and (A.24) and (A.25). 

We will now show how this procedure works in detail.
Consider the following Bianchi identity obtained in
\cite{Andrianopoli:1997cm}
\be{cuv1}
\nabla^2 z^i  = g \left( F^\Lambda - \bar{L}^\Lambda \bar{\psi}_A
\wedge \psi_B \epsilon^{AB} - \bar{L}^\Lambda \bar{\psi}^A \wedge
\psi^B \right)k^i_\Lambda(z)
\ee
where $z^i$ is thought of as zero form in ${\cal N}=2$ superspace. 
Expanding the left hand side in terms of flat space derivatives we
obtain
\be{flexp1}
\nabla^2 z^i = E^M E^N {\cal D}_N {\cal D}_M z^i + T^M {\cal D}_M z^i
\ee
The first term in the right hand side of the above equation is the
commutator of the covariant derivatives. We need  to look for
indices with either $M$ or $N$ being the equal to the one spatial
index and one anti-chiral spinorial index in the ${\cal N}=1$ superspace, 
to obtain the commutator of covariant
derivative in \eq{dext}. 
Now consider the torsion term in \eq{flexp1}, from
\cite{Andrianopoli:1997cm} equation (A.23),  
we see that a torsion with bosonic
component vanishes, therefore $M$ must take values along the spinorial
directions, this allows us to write the torsion contribution as
\be{wrtor}
T^M {\cal D}_M z^i = \rho_A {\cal D}^A z^i + \rho^A{\cal D}_A z^i
\ee 
From the equation 
\be{covderz}
\nabla z^i = \bar{Z}^i_a V^a + \bar{\lambda}^{iA} \psi_A
\ee
we see that the second term of \eq{wrtor} drops out,
as the covariant derivative of $z^i$ does not have any component in
the anti-chiral directions. 
The component of the torsion relevant for us is
\be{soltor}
\rho_A = 
\left[ i g S_{AB} \eta_{ab} + \epsilon_{AB} ( T^-_{ab} + U_{ab}^+ )
\right] \gamma^b \psi^B \wedge V^a + \ldots, 
\ee
Now we look for the components which contribute from the curvature on
the right hand side of \eq{cuv1}, 
again the component relevant for us is
\be{curvf}
F^\Lambda = 
\left(
i f^\Lambda_i \bar{\lambda}^{iA} \gamma_a \psi^B \epsilon_{AB} 
\right)\wedge V^a  + \ldots 
\ee
Equating the coefficients with one anti-chiral spinor index and a
spatial index of \eq{cuv1} and \eq{flexp1}, we obtain
\be{eqbo}
ig f_j^\Lambda \bar{\lambda}^{jA} \gamma_a 
\epsilon_{AB} k^i_\Lambda = \bar{\lambda}^{iA} 
\left(i g S_{AB}\eta_{ab} + \epsilon_{AB}(
T^{-}_{ab} + U_{ab}^+ )\gamma^b  \right) + 
[{\cal D}_B, {\cal D}_a] z^i
\ee
To extract the commutator in \eq{dext} we need to consider the the
anti-chiral spinor index to be that which corresponds to the unbroken
${\cal N} =1$. From the breaking pattern discussed in 
section 3, this supersymmetry corresponds to $B=1$. Substituting the fields,
including the ${\cal N}=2$ gravity backgrounds 
and the scalings involved in obtaining the rigid limit, we obtain
\be{prering1}
T_{ab}^- \bar{\lambda}^{i2} = i \bar{\lambda}^{j2}
\gamma_a f^i_{jl} z^l
\ee
As we have seen from the calculation of the mass  of the fermions in
the spontaneously broken theory in \eq{mass}, the gauginos of the
${\cal N}=1$ theory correspond to $\lambda^{i1}$ which we denote by
$\lambda^i$. Thus the above equation reduces, after a conjugation
and  some gamma matrix manipulation to 
\be{compring1}
i f^i_{jk} \lambda_{\alpha}^j z^k  = \frac{1}{2}
T^-_{\alpha\beta} \lambda^{\beta i}
\ee
or in matrix notation to
\be{finring1}
[\lambda_\alpha, z] =  
\frac{1}{ 2} T_{\alpha\beta}
\lambda^\beta
\ee
where 
\footnote{The factor 2 occurs in passing from the conventions of the
Lorentz generators of \cite{Andrianopoli:1997cm} to that of 
\cite{Wess:1992cp}. 
We use the conventions of the latter to write the
chiral ring relations in their final form.}
$T_{\alpha\beta} =  2\gamma^{ab}_{\alpha\beta} T^-_{ab}$
and  $T^-$ stands for the self-dual components of the graviphoton
field strength 
background, we have suppressed the superscript $^{(b)}$ denoting the
background for convenience.

There are two methods to obtain the second chiral ring relation. We
will present the quick  
method here to have an understanding of the structure of the second
ring relation 
and then present the detailed
method using the Bianchi identities.  
Thinking of the relation in \eq{finring1} as a relation in ${\cal
N}=2$ superspace, 
we can obtain the 
second ring relation by 
performing a supersymmetry transformation with the broken
supersymmetry generators, that is generators with the $2$ index. 
Thus the second ring relation is given by
\be{qrin2}
if^i_{jk} \lambda_\alpha^j \lambda_\beta^k  =  
\frac{1}{2} T_{\alpha\beta} D \delta^{in}  + 
\frac{1}{2} \psi_{\alpha\beta\gamma} \lambda^{\gamma i} 
\ee
where $D\delta^{in}$ is the expectation value of the $D$-term that we have
turned on due to gauging from the hypermultiplet \footnote{It is the
22 component in the supersymmetry variation \eq{redshg}.}. We will
see that in the detailed analysis $D= 2\Lambda^2 m$ 
and $\psi_{\alpha\beta\gamma}$ is the ${\cal N}=1$ gravitino field 
strength background.  To obtain \eq{qrin2} from \eq{compring1} we 
we have used the following supersymmetry transformations
\be{broksu}
\delta_2 z^i = \lambda_\alpha^i \eta^\alpha,  \;\;\;\;
\delta_2 \lambda^{\alpha i}  = D\delta^{in} \eta^\alpha, \;\;\;\;
\nonumber
\delta_2  T_{\alpha\beta} =
-  \psi_{\alpha\beta\gamma} \eta^\gamma.
\ee

Let us now see in detail how the second ring relation comes about
using Bianchi identities. 
Consider the Bianchi identity which involves $\nabla^2$ acting on 
$\lambda^{iA}$
\be{cuv2}
\nabla^2 \lambda^{iA} + \frac{1}{4} \gamma_{ab} R^{ab} \lambda^{iA} +
\frac{i}{2} \hat{K} \lambda^{iA} + \hat{R}^i_j \lambda^{jA} -
\frac{i}{2} \hat{R}^A_B \wedge \lambda^{iB} =0
\ee
Again following the similar procedure of 
expanding the left hand side of the above equation in flat
space indices and looking for the components with one
anti-chiral spinorial index and one spatial index, we obtain the
following equation
\bea{eqincomp}
\frac{-i}{4} \eta_{ca} \left( 2 \gamma^{[d} \rho_B^{b] c} + \gamma^c
\rho_B^{db} \right) (\gamma_{db} \lambda^{iA} )
- i \epsilon_{CB}\gamma_a \lambda^{kC}f^i_{jk}
  \lambda^{jA} \\ \nonumber
  + \sqrt{2}\epsilon_{CB}T^-_{ab}\gamma^b  W^{iAC}  m_p \Lambda
+ [D_B, D_a] \lambda^{iA} =0.
\eea
Here we have substituted
for all the fields and their scalings involved, except $W^{iAC}$. 
Now consider the   component  with $B=1$ and $A=2$. After some gamma
matrix manipulation and using the on shell conditions on the
gravitino background we obtain  the following chiral ring relation.
\be{compring2}
i  T_{\alpha\beta} \Lambda^2  m \delta^{in} 
+ \frac{1}{ 2}  \psi_{\alpha\beta\gamma} \lambda^{\gamma i}= i f^i_{jk}
\lambda_\alpha^j \lambda_\beta^k
\ee
In the above equation 
we have substituted the value of $W^{iAC}$ at the ${\cal N}=1$
vacuum, we will show subsequently how fluctuations around this vacuum
value are $\bar{D}$ exact terms. The gravitino field strength
$i\rho^{ab}_1$ has been replaced by $\psi^{ab}$ \footnote{This factor of $i$
occurs when changing from the signature conventions of
\cite{Andrianopoli:1997cm} to \cite{Wess:1992cp}. }.

It is important to note that the for the component $i=n$, the $U(1)$
part  the right hand side of \eq{compring2} vanishes and we are left with the
relation 
\be{subring}
i T_{\alpha\beta} \Lambda^2 m + \frac{1}{2} \psi_{\alpha\beta\gamma}
\lambda^{\gamma n} =0
\ee
This is in contrast with the proposal of \cite{Ooguri:2003qp} and
\cite{Ooguri:2003tt}, there it was assumed that the presence of the
graviphoton renders the gaugino non-Grassmanian, thus according to
this proposal there would be a contribution from the 
the Abelian part on the right
hand side of \eq{compring2}. We  see here that  within the framework of
${\cal N}=1$ theories obtained from spontaneously broken ${\cal N}=2$
theories, the presence of the graviphoton in the chiral ring relations
is a consequence of  Bianchi identities and traditional supergravity
tensor calculus. As a result there is no drastic change in the Grassmann
nature of the gaugino superfield or of the superspace coordinates.

As the two chiral ring relations 
in \eq{compring1}
and \eq{compring2} are written in terms of bottom components of
${\cal N}=1$ superfields, we can promote these equations to an 
equation in ${\cal N}=1$ superspace. Thus the chiral ring relations 
are given by
\bea{fring}
[W_\alpha, \Phi] &=& -\frac{i}{2\Lambda^2 \tilde m} F_{\alpha\beta}
W^\beta, \\ \nonumber
[W_\alpha, W_\beta] &=& F_{\alpha\beta} + 2 G_{\alpha\beta\gamma}
W^\gamma
\eea
Here we have used the fact that the bottom component of
$G_{\alpha\beta\gamma}$ is given by $ \psi_{\alpha\beta\gamma}/4$
\cite{David:2003ke}. We have also rescaled the graviphoton with
$F_{\alpha\beta} = m \Lambda^2/ \sqrt{N}
T_{\alpha\beta}$. This ensures that the  dimensions for the
graviphoton is $3$. The factor of $\sqrt{N}$ appears in writing the
\eq{compring2} as a matrix equation since the $U(1)$ generator is
$I/\sqrt{N}$, finally $m =\sqrt{N} \tilde {m}$. 

In deriving the ring equations \eq{fring} from the Bianchi
identities, we have neglected the fluctuations in the $D$ terms and
just substituted the background $U(1)$ valued D-term proportional to 
$\Lambda^2 m$. We argue that the fluctuations to the D-terms arising
from the expansion around the ${\cal N}=1$ vacuum are $\bar{D}$ exact
terms and thus vanish in the chiral ring. 
Indeed ${\cal N}=2$ vector multiplet can be organized into a
constrained chrial ${\cal N}=2$ superfield  as follows
\be{bigsup}
\Psi = \Phi + \theta^2 W + (\theta^2)^2 D
\ee
Here the auxiliary $D$ is an ${\cal N}=1$ superfield whose bottom
component gets an expectation value at the vacuum.
The reality constraint on the ${\cal N}=2$ superfield is  given by
\be{bigconst}
(\epsilon^{AB} D_A \sigma^{\mu\nu} D_B ) ^2 \Psi = - 96 \Box
\Psi^\dagger
\ee
Acting by $(\bar{D}_1)^2$ on both sides of the equation we obtain
\bea{dconst}
4\Box (D_2)^2 \Psi &=& 3 (\bar{D}_1)^2 \Box \Psi^\dagger, \\ \nonumber
4 \Box D &=& 3 (\bar{D}_1)^2 \Box\Phi^\dagger
\eea
Thus $D$ is $\bar{D}$ exact, but
a  constant expectation value of $D$ is not subject to this
constraint, due to the presence of the d'Alembertian on both sides of
the equation.
Thus fluctuations in the $D$ term can be neglected in the second chiral ring 
equation of \eq{fring}.

The chiral ring relations for the hypermultiplets do not receive any
gravitational corrections. This can be seen by careful
considerations of the Bianchi identities involving the hypermultiplets from
\cite{Andrianopoli:1997cm} and the scalings given in
(\ref{scalnewhy}), but here we present a quick argument.
Consider the commutator in \eq{dext} acting on the chiral multiplet 
$Q_1$ obtained from the hypermultiplet, 
there is only one contribution to the curvature term, due to
the presence of the gauge index on $Q_1$. As $Q_1$ is a scalar it
does not receive any contribution from the gravitino field strength
because the  chiral field in \eq{dext} must carry at least one
spinor index to receive a contribution. 
From the  previous examples  we have seen that 
the torsion term in \eq{dext} always
comes with a covariant derivative in the $2$ direction, as a ${\cal
N}=2$ superfield. 
If $D_2 Q_1$ is non zero, it should  be proportional to $Q_2$ as 
$D_2$ mixes the two chiral multiplets of hypers. But $D_2Q_1$ 
cannot contain any chiral components since
$Q_2$ is in the anti-fundamental representation, while $D_2Q_1$  must
be in the fundamental representation. Therefore, from this analysis we
see that the only contribution to the commutator in \eq{dext} is from
the curvature term and that is because of the gauge index on $Q_1$.
Thus the chiral ring relations for the hypermultiplets are given by
\be{ringhyp}
Q_1 W_\alpha =0, \;\;\;\; W_\alpha Q_2 =0
\ee
This completes our analysis of the gravitational deformed chiral rings
using the partially broken ${\cal N}=2$ theory.

\section{Generalized Konishi Anomaly equations}

In ref.\cite{Alday:2003ms}, we had obtained gravitational corrections to the
effective superpotential using generalized Konishi anomaly
equations. In that derivation we had assumed that the ${\cal{N}}=1$
gauge function and K\"{a}hler functions at the classical level were
trivial and only the classical superpotential was an arbitrary
single trace polynomial in the chiral field. Furthermore the chiral
ring relation that was used, although similar to eqs.\eq{fring}, was not
exactly the same. Specifically the ring relations used there were
\begin{eqnarray}
[W_{\alpha}, \Phi] &=& 0 \\
{[}W_{\alpha}, W_{\beta}{]}  &=&  F_{\alpha \beta} + 2 G_{\alpha \beta
\gamma} W^{\gamma}
\label{oring}
\end{eqnarray}
So while the second ring relation is the same as in eq.\eq{fring}, the first
one differs. Note that the right hand side of the first equation
\eq{fring},
scales as $1/\Lambda^2$. Therefore in the limit of the supersymmetry
breaking
scale $\Lambda \rightarrow \infty$, the ring relations derived here
reduce to the ones of ref.\cite{Alday:2003ms}. Moreover, in this limit, the gauge
function (\ref{gaugfns}) and K\"{a}hler metric become field
independent and therefore 
in this limit we reproduce the situation considered in ref.\cite{Alday:2003ms}.

In this section we will analyze the modifications in the anomaly
equations when $\Lambda$ is finite and show that there
are $\Lambda$-dependent
corrections. We will also see that the consistency
of the anomaly equations in the presence of graviphoton field strength 
crucially requires the ${\cal{N}}=2$ relation between the
superpotential and the
gauge function, namely that they come from the prepotential.

Since $W_{\alpha}$ and $\Phi$ do not commute in the chiral ring
\eq{fring},
at first sight it appears that the 
derivation of generalized Konishi anomaly equations become very
difficult as now the ordering of $\Phi$ with respect to $W_{\alpha}$
inside traces cannot be ignored. However by a field redefinition ring
relations of \eq{fring} can be transformed into eq.(\ref{oring}). Indeed  
from the second ring relation we can obtain the following relation
\begin{equation}
[W_{\alpha}, W^2 ] = -2 F_{\alpha \beta} W^{\beta}
\end{equation}
Using this fact it is easy to see that by a field redefinition
\begin{equation}
\Phi \rightarrow \Phi + i\frac{1}{4\tilde{m} \Lambda^2} W^2
\label{newphi}
\end{equation}
the first ring relation of eq \eq{fring} goes over to that of
eq.(\ref{oring}). By substituting the above field redefinition in the action we
can read off the gauge function and the superpotential in  terms of
the redefined chiral field $\Phi$. In Appendix B, it is shown that the 
superpotential remains unchanged while the gauge function becomes 
$\Phi$ independent constant modulo $\bar{D}$-terms, and all the $\Lambda$
dependence drops out. 
As this point is important we will demonstrate this cancellation for
the first $\Lambda$ dependent term here.
This $\Lambda$ dependent term arises from the cubic term of the
prepotential in \eq{prepot1}. We see from \eq{gaugfns} the
contribution of the cubic term to the kinetic term for the gauginos
is 
\be{cukin}
-\frac{ g_1 }{ \Lambda^2 } \frac{i}{4} {\rm Tr} (\Phi W^2)
\ee
Now performing the shift  of \eq{newphi} in the superpotential on the
term $\tilde{m} \frac{g_1}{2} \Phi^2$one
gets the same contribution as in \eq{cukin} but  with the opposite
sign, thus this $\Lambda$ dependent term cancels.
It is important to stress here that this 
cancellation between the gauge kinetic terms coming from the original
gauge function and the superpotential, crucially depends on the 
$N=2$ relation between the two. Had we taken an arbitrary gauge
function, which is certainly allowed in $N=1$ theory, this
cancellation would not have taken place (in terms of the redefined 
fields (\ref{newphi})) and there would have been a surviving non-trivial gauge 
function. We will show below that the resulting generalized Konishi
anomaly equations then would have been inconsistent in the presence of
graviphoton field strength. 

The K\"{a}hler function, i.e. the kinetic term for the chiral fields, is
however still a non-trivial function of redefined chiral fields for finite
$\Lambda$. Note however that all these non-trivial $\Phi$ dependent
terms are non-renormalizable. One might wonder if under a transformation $\Phi 
\rightarrow \Phi + f(\Phi)$ where $f$ is an arbitrary function,
needed to derive generalized Konishi anomaly equations, the
non-trivial K\"{a}hler function would contribute. Since our equations are
chiral ring equations (i.e. modulo $\bar{D}$ exact terms), the K\"{a}hler
function can contribute only through generalized Konishi anomaly.  
We do not expect that the non-renormalizable terms would contribute to
the anomaly,
therefore the structure of anomaly contribution to the Schwinger-Dyson 
equations resulting from the above transformation would be governed by
the trivial renormalizable (field independent) kinetic term. 
 
It then follows that the equations derived in ref.\cite{Alday:2003ms} 
remain unchanged
and the effective superpotential receives no correction for finite
$\Lambda$. However now we would like to show that if there had been a
non-trivial gauge function (after the field redefinition (\ref{newphi})), we
would have obtained inconsistent equations in the presence of
graviphoton field strength. In the following $\Phi$ will always denote
the redefined field (\ref{newphi}) so that in the chiral ring it commutes with
$W_{\alpha}$. Recall that under an infinitesimal transformation
$\delta \Phi = f(\Phi, W)$, the generalized Konishi anomaly is given by
\begin{equation}
\frac{1}{32\pi^2}\frac{\delta f_{ji}}{\delta \Phi_{k \ell}} A_{ij,k \ell},
\end{equation}
with
\begin{equation}
A_{ij,k \ell} = (W^2)_{kj} \delta_{i\ell} + \delta_{kj} (W^2)_{i\ell} - 2
W^{\alpha}_{kj} W_{\alpha i\ell} + \frac{1}{3} G^2 \delta_{kj} \delta_{i\ell}.
\label{an}
\end{equation}
where the last term is the gravitational contribution to the anomaly.
Now consider a transformation
\begin{equation}
\delta \Phi = f(\Phi,W)= F_{\alpha \beta} \frac{1}{z-\Phi} + 2 G_{\alpha \beta
\gamma} \frac{W^{\gamma}}{z-\Phi}
\end{equation}
The corresponding generalized Konishi anomaly vanishes (modulo
D-terms). Indeed by using the
anomaly (\ref{an}) and the chiral ring relation (\ref{oring}) we find that the
anomaly is proportional to:
\begin{eqnarray}
&~& 2 {\rm Tr}(\frac{W^2}{z-\Phi})(  F_{\alpha \beta} {\rm
Tr}(\frac{1}{z-\Phi}) + 2 G_{\alpha \beta \gamma} {\rm
Tr}(\frac{W^{\gamma}}{z-\Phi}))\\   \\ \nonumber
&=&2 {\rm Tr}(\frac{W^2}{z-\Phi}) {\rm Tr} (\frac{\{ W_{\alpha}
  W_{\beta} \} }{z-\Phi}) =0
\end{eqnarray}
where we have used the fact that in the chiral ring $G^2$ and
$(F.G)_{\gamma}$ multiplied by $F$ or $G$ vanish (see
eqs.(2.13)-(2.20) of ref.\cite{Alday:2003ms} for proof).  

Under this transformation the change in classical superpotential
vanishes:
\begin{equation} 
\delta {\cal W}(\Phi) =   F_{\alpha \beta} {\rm Tr}(\frac{{\cal
W}'}{z-\Phi}) + 2 G_{\alpha \beta \gamma} {\rm Tr}({\cal W}'\frac{W^{\gamma}}
{z-\Phi}) ={\rm Tr} ({\cal W}'\frac{\{ W_{\alpha} W_{\beta} \} }{z-\Phi}) =0
\end{equation}
Here and in the following prime denotes derivative with respect to $\Phi$.
If there is a non-trivial gauge function ${\rm Tr}\tau(\Phi)
W^2$ in the
action, then the corresponding change is
\begin{equation}
\delta ({\rm Tr}\tau W^2)=   F_{\alpha \beta} {\rm
Tr}(\tau'\frac{W^2}{z-\Phi}) + 2 G_{\alpha \beta \gamma} {\rm
    Tr}(\tau'\frac{W^{\gamma} W^2}{z-\Phi})=  F_{\alpha \beta} {\rm
Tr}(\tau'\frac{W^2}{z-\Phi}) 
\end{equation}
where we have used the fact that
\begin{eqnarray}
  {\rm Tr}(\tau'\frac{W^{\gamma} W^2}{z-\Phi}))&=&  \frac{1}{2}{\rm
    Tr}(\tau'\frac{\{W^{\gamma}, W^2\} }{z-\Phi})\\ &=& -\frac{1}{3}(G^2{\rm
    Tr}(\tau'\frac{W^{\gamma} }{z-\Phi}) +(G.F)^{\gamma} {\rm
    Tr}(\tau'\frac{1}{z-\Phi})) 
\end{eqnarray}
and the fact that $G^2$ and $(G.F)^{\gamma}$ when multiplied by
another $G$ vanishes in chiral ring.
In the second equality above, we have repeatedly made use of the
second ring relation (\ref{oring}).

This shows that the change in the gauge kinetic term does not vanish
if the gauge function $\tau$ is a non-trivial function of $\Phi$ (the
only exception for single trace gauge function is if
$\tau'$ is proportional to $G^2$). Thus for a non-trivial gauge function
the resulting Schwinger-Dyson equations would be inconsistent with the
chiral ring relations (\ref{oring}). As shown in the Appendix B, the triviality
of the gauge function is a consequence of the precise ${\cal N}=2$
relation between the superpotential and the gauge function, which
arises due to the fact that the ${\cal N}=1$ theory is obtained by
partial spontaneous breaking of ${\cal N}=2$ theory.

\section{Conclusions}

In this paper we have discussed the ${\cal N}=1$
effective field theory arising in the rigid 
limit of a partially broken ${\cal N}=2$
supergravity coupled to vector- and hyper-multiplets,
in the presence of a non-trivial background
graviphoton $F_{\alpha\beta}$ and gravitational  
superfield $G_{\alpha\beta\gamma}$. We have shown that
the effective field theory one obtains is the one
discussed in \cite{Dijkgraaf:2002dh} and subsequent literature. Moreover,
we have also derived the chiral ring relations  
in the presence of the above backgrounds. We have shown that
they are precisely those used in the literature \cite{
Ooguri:2003qp,Ooguri:2003tt,David:2003ke,Alday:2003ms}
to connect the generalized gravitational superpotentials
of the SYM theory to the coefficients
of the topological expansion in the 
corresponding matrix model integral. We should stress 
however that in our approach the chiral ring is a consequence of the 
Bianchi identities of the underlying ${\cal N}=2$ theory and therefore
no drastic change in the
Grassmann nature of the gaugino superfield or of the
superspace coordinates is postulated.

Another important result of our analysis, supporting the
${\cal N}=2$ interpretation, is that, in the presence
of $F_{\alpha\beta}$, consistency of the loop
equations requires that a non trivial 
gauge function, in principle arbitrary
for an ${\cal N}=1$ SYM theory,  should in fact 
be related to the superpotential 
precisely in the way given by  ${\cal N}=2$ supersymmetry.

A point which deserves better understanding is whether
our effective field theory can be derived from the
effective action of wrapped  D5-branes. Whereas the
relation of fluxes to gauging seems to be well
understood on the closed string side of the
closed/open string duality, on the D5-branes side 
this is not completely clear. We recall that the crucial 
ingredient for this partial breaking is the gauging of two
translational symmetries in the hypermultiplet space; with
the two $U(1)$ gauge fields being the graviphoton and another
gauge field that decouple completely in the rigid limit. 
Here we indicate possible mechanisms for this gauging on the 
D-brane side.
The
obvious candidate for the latter is the center of mass $U(1)$
of the D-brane system. The hypermultiplets that can naturally play a
role in this problem are the universal one (associated to the dilaton)
and the one associated with the (1,1) form dual to the 2-cycle
which is wrapped by the D-5 brane. Consider the Chern-Simons term
on the D-brane world-volume
\begin{equation}
S_{cs} \sim \int d^6 x C_{4} \wedge {\rm Tr} F
\end{equation}
where $F$ is the world-volume gauge field strength and $C_{4}$ is the
RR 4-form potential. Writing $C_{4}= T_{2} \wedge \omega_{(1,1)}$
where
$\omega_{(1,1)}$ is the (1,1)-form dual to the 2-cycle wrapped by the
D-brane and $T_2$ is a 2-form in the remaining directions, and dualizing
$T_2$ in the non-compact 4-dimensional part of the world volume, we
find a coupling of the form ${\rm Tr} A^{\mu} \partial_{\mu} b$ where
$b$ is the scalar field dual to $T_2$. Thus we see that the center of
mass $U(1)$ gauges the translational symmetry of the scalar field $b$.
Note also that writing the gauge field as $A = A_{cm} {\bf 1}/\sqrt{N}
+A^i T_i$ where $T_i$ are the $SU(N)$ generators and the factor of
$\sqrt{N}$ is included so that the kinetic term for $A_{cm}$ is
normalized to unity, we find that the Chern-Simons coupling term is
$\sqrt{N} A_{cm}^{\mu}\partial_{\mu}b$. This is in accordance with the 
scaling of
charge $m$ as $\sqrt{N} \tilde{m}$ observed in the comments following
eq.(\ref{fring}).

The issue of the second $U(1)$ gauging by graviphoton is less clear in
the context of D-branes. In IIB  theory under consideration, the
graviphoton appears from the RR 5-form field strength $F_{5}= F_{2}^{gr}\wedge
\Omega$ where $\Omega$ is the holomorphic 3-form on the (non-compact)
Calabi-Yau space. Given the fact that in the presence of D5- branes
there is a non-zero flux of the associated RR field strength $F_{3}$ 
through the non-compact Calabi-Yau space, the 10-dimensional
Chern-Simons term $\int d^{10} x F_{5} \wedge F_{3} \wedge B_{2}$
could in principle gauge the translational symmetry of the scalar
field dual to the NS-NS 2-form potential $B_{2}$. The charge then
would be proportional to the integral $\int_{CY} \Omega \wedge F_{3}$
over the Calabi-Yau space. 
Thinking of the non-compact Calabi-Yau as a $K_3$ fibered over a
plane, the holomorphic 3-form can be written as $\Omega = \omega
\wedge dz$,
where $\omega$ is the holomorphic 2-form inside $K_3$ and $dz$ is the
holomorphic 1-form on the complex plane, thus $\Omega = d\alpha$
where 
\be{holform}
\alpha =  \int_{z = z_0 }^{z}  \omega \wedge dz,
\ee
therefore 
\be{extformman}
\int_{CY} \Omega \wedge F_{3} 
= \int \alpha \wedge d F_3   + 
\int \alpha \wedge F_3|_\infty
\ee
here by $\infty$ we mean the boundary on the complex plane at infinity. 
The first term in the above equation is finite, infact  $dF_3$ is
localized at the position of the brane. However one needs to put a
cut off to give a meaning to the second term.
Thus it is not clear if 
the theory on D5-brane wrapped on a 2 cycle of a non-compact
Calabi-Yau provides a realization of the spontaneously broken theory
considered in this paper. It will be interesting to study this issue
further.

\acknowledgments
This research  was supported in part by EEC contract EC
HPRN-CT-2000-00148.

\appendix

\section{Conventions}

We list here the conventions for the flat hyper K\"{a}hler manifold.
The manifold is parametrized by the coordinates $q = (q^{Au},
\bar{q}^{A\bar{u}} )$, 
where $A =1,2$ and $u = 1, \ldots m$. They form a set
of $2m$ complex coordinates. The metric on this manifold is given by
\be{methyp}
ds^2 = dq^{Au}dq^{B\bar{u}} \delta_{AB}\delta_{u\bar{u}} 
\ee
We will denote this metric as $h_{\rm{u} \rm{v}}$, here $\rm{u},
\rm{v}$ refer to the entire set $2m$
holomorphic and the anti-holomorphic coordinates, 
thus $h_{AuB\bar{u}} = \frac{1}{2} \delta_{AB} \delta_{u\bar{u}}$.
The vielbein one form is  given by
\be{viel}
{\cal U}^{Aai} = {\cal U}^{Aai}_{Bu} dq^{Bu} + {\cal
U}^{Aai}_{B\bar{u}} d\bar{q}^{B\bar{u}}
\ee
Here we have split the $Sp(2m)$ indices as $ai$ with $a = 1,2$ and $i
=1, \ldots m$. The explicit form for the vielbein are
\be{solviel}
{\cal U}^{Aai}_{Bu} = \frac{1}{\sqrt{2}} \delta^{a1} \epsilon^{AB} \delta^{iu},
\;\;\;\;\;
{\cal U}^{Aai}_{B\bar{u}} = \frac{1}{\sqrt{2}} 
\delta^{a2} \delta^{AB} \delta^{iu},
\ee
The vielbein defined this way
satisfy the following required conditions for the hyper-K\"{a}hler
manifold
\bea{condviel}
h_{\rm{u} \rm{v}} &=& 
{\cal U}^{Cai}_{\rm{u} } {\cal U}^{Dbj}_{\rm{v}}
\epsilon_{CD}\epsilon_{ab}\delta{ij}, 
\\ \nonumber
({\cal U}^{Aai})^* &=& \epsilon_{AB} \epsilon_{ab}\delta_{ij}
{\cal U}^{Bbj}, \\ \nonumber
h_{\rm{u} \rm{v} } \epsilon^{AB} &=&
( {\cal U}^{Aai}_{\rm u}  {\cal U}^{Bbi}_{\rm v} + 
{\cal U}^{Aai}_{\rm v}  {\cal U}^{Bbi}_{\rm u} )
\epsilon_{ab}\delta_{ij} 
\eea

\section{ $\Lambda$ independence of the gauge kinetic term
}

In this appendix we show that after the field redefinition \eq{newphi}
the $\Lambda$ dependence of the gauge kinetic terms drops out modulo
$\bar{D}$ terms for the ${\cal N}=1$ theory obtained from partially broken
${\cal N}=2$. 
We will need the following identities in the chiral ring which are
obtained by simple algebraic manipulations from the basic ring
relations \footnote{For a proof of these identities see
\cite{Alday:2003ms}.}.
\bea{iden1}
[W_\alpha, W^2] &=& -2 F_{\alpha\beta} W^\beta,  \\ \nonumber
\{W_\alpha, W^2 \} &=& - \frac{2}{3} ( G^2 W_\alpha +
G_{\alpha\beta\gamma} F^{\beta\gamma} ).
\eea
To simplify the first ring relation in \eq{fring} we perform the
following field redefinition 
\be{baredef}
\Phi \rightarrow \Phi +  \alpha  W^2, \;\;\;\; \alpha =
\frac{i}{4\tilde{m} \Lambda^2}
\ee
The redefined $\Phi$ now commutes with $W_\alpha$, but we have
to perform the same shift in the action of the spontaneously broken
theory. As we have seen in section 3.  
after breaking ${\cal N}=2$ symmetry to ${\cal N}=1$ we
following obtain the superpotential
\be{suppot}
{\cal W} = \tilde{m} \sum_{k=1}^n \frac{g_k}{k+1} {\rm Tr} \Phi^{k+1}
\ee
Apart from the trivial $\Lambda$ dependent term in the gauge kinetic
function this theory also has a non trivial gauge function obtained from the
prepotential
\be{prepot}
{\cal F}' = 
\frac{1}{\Lambda^2} \sum_{k=1}^n \frac{g_k}{(k+1)(k+2)} {\rm Tr} \Psi
^{k+2}
\ee
A convenient way of writing the gauge function of the ${\cal
N}=1$ theory is  $i \int d^2 \hat{\theta} {\cal F}'(\Psi) $ 
and  substituting $\Psi = \Phi +
\hat\theta W_\alpha + \hat{\theta}^2 D$ in the prepotential 
and extracting out the coefficient
of $W^2$ \footnote{This definition of the gauge function gives the
correct normalization for the usual kinetic term for the prepotential
$\Psi^2/2$ which is $-i/4$}. 
Roughly this is given by the second derivative of the
prepotential \eq{prepot}. However as $\Phi$ and $W_\alpha$ don't
commute with each other we have to be careful of the orderings in
higher powers of $\Psi$. Our aim now is to make the redefinition of 
\eq{baredef} in both the gauge function and the superpotential 
and simplify the resulting higher powers of $W^2$ using
the ring relations of \eq{iden1}. 
We claim that  after the redefinition to the 
chiral multiplet which commutes with 
$W^\alpha$ the  $\Lambda$ dependence of the gauge kinetic term drops
out.

As a first test we look at the terms which are of the type 
${\rm Tr}(\Phi W^2)$. There are two contributions, one from the
superpotential which is given by 
$\tilde{m} \alpha g_1 {\rm Tr}(\Phi W^2)$, the contribution from
the gauge function is given by $-i\frac{1}{4\Lambda^2 }  {\rm Tr} (\Phi
W^2)$ and thus for the value of $\alpha$ given in 
\eq{baredef} they cancel. The $-\frac{1}{4}$ factor comes if one keeps
track of the $\hat{\theta}$ orderings.
Terms proportional to $W^4$ can be
rewritten due to the following identity
\bea{ident2}
{\rm Tr } ( W^2 W^2 \Phi^k) &=& 
\frac{1}{2}{ \rm Tr} \{ W^\alpha { W_\alpha, W^2} \Phi^k \}, \\
\nonumber
&=& -\frac{1}{3} {\rm Tr} ( G^2 W^2 \Phi^k + ( G\cdot F)^\alpha
W_\alpha \Phi ^k ) 
\eea
We also note that
\be{ident3}
[W_\alpha, W^4] =0 
\ee
From \eq{ident2} and \eq{ident3}  
it is easy to see that contribution proportional $(W^4)^p$ for $p>1$
vanishes in the chiral ring. This is seen as follows
\bea{ident4}
{\rm Tr} ( W^4 (W^4)^{p-1} \Phi^k ) &=& 
\frac{1}{2}{\rm Tr} ( W^\alpha{ W_\alpha, W^2} (W^4)^{p-1} \Phi^k ) , 
\\ \nonumber
&=& - \frac{1}{3} {\rm Tr} \left( G^2 W^2 + (G\cdot F)^\alpha W_\alpha )
(W^4)^{p-1} \Phi ^k  \right)
\eea
For dimensional reasons any reduction of powers of $W$ in the last
equation of \eq{ident4} would involve higher powers of $F$ or $G$. Any
multiplication of $F$ or $G$ with $G^2$ or $G\cdot F$ vanishes in the
chiral ring \cite{Alday:2003ms}. 
Therefore contributions proportional to $(W^4)^p$
vanishes for $p>1$.

We are now left with terms proportional to $(W^2)^{2p +1}$. We
will now show that they too vanish. 
Consider the contribution to $(W^2)^{2p +1}$ from terms proportional
to $g_k$ for sufficiently large $k$ both from the 
superpotential and from the gauge kinetic term. 
The superpotential contributes the following term.
\be{consup}
{\cal W}_{2p+1} =  \tilde{m} \alpha^{2p +1} 
\frac{g_k}{k+1} \frac{(k+1)!}{(2p +1)! (k -2p) !} 
{\rm Tr} ( (W^2 )^{2p +1} \Phi^{k -2p} )  
\ee
We will denote the combinatorial factor in front by $N_s = {}^{k+1}
C_{2p +1}$. 
From the gauge kinetic term the contribution to $(W^2)^{2p +1}$ can be
organized into two kinds. One kind of terms can be manipulated due to
the cyclic properties of the trace to a term proportional to 
${\rm Tr } ( (W^2)^{2p +1} \Phi^{k-2p} )$, 
the second kind of terms 
can be manipulated by the cyclic properties of the trace and
\eq{ident3} to a term proportional to 
${\rm Tr } ( (W^2)^{2p-1} W^{\alpha} W^2 W_{\alpha} \Phi^{k-2p} )$.  
We will now find the precise combinatorial coefficient in front of
this term. The second type of  terms arise from first choosing 
$k-2p$ $\Phi$'s out of
$k+2$ terms and then looking for various arrangements of the $W$'s
such that there are odd number of $W^2$ between two single $W^\alpha$'s. 
This is because if there is an even number of
$W^2$ between the two $W_{\alpha}$ they can be commuted through 
so that the two $W_{\alpha}$ become adjacent giving rise to $W^2$,
and therefore gives a term of the first type.
The number of ways of having odd number of $W^2$'s between two
$W_{\alpha}$'s is given by
$p(p+1)$. This is argued as follows; from the remaining $2(p+1)$
$\Phi$ we need to take two $W_{\alpha}$ and the remaining $2p$
$W^2$. In
doing so we must have odd number of $W^2$ in between the positions of
the two $W_{\alpha}$. To
count the number of ways we can have odd number of $W^2$ between the
two $W_{\alpha}$, let $n$ denote the position of the second
$W_{\alpha}$. Clearly $n=3,...,2p+2$. Then the position of the first 
$W_{\alpha}$ must be $n-2k$ for some integer $k$ so that the number of
$W^2$ in the middle is $2k-1$. Clearly $k=1,...[\frac{n-1}{2}]$ where
the square bracket means the integer part of the argument. The number
of possible values of $k$ therefore is $(n-1)/2$ for odd $n$ and
$(n-2)/2$ for even $n$. Writing $n=2m+1$ for odd $n$ and $2m+2$ for
even $n$ we see that the range of $m$ in either case is from $1$ to
$p$ and the number of possible $k$ is $m$. Thus the total number of
such terms is $2\sum_{m=1}^p m = p(p+1)$. All these terms, by using
the cyclicity of the trace and by moving even powers of $W^2$ across
$W_{\alpha}$ can be brought to the form ${\rm Tr}((W^2)^{2p-1}W^{\alpha}
W^2 W_{\alpha} \Phi^{k-2p})$. 
Thus the total number of terms of the second type    
is $N_{o} = {}^{k+2}C_{k-2p} p (p+1)$, where the binomial factor comes
from possible ways of picking $k-2p$ $\Phi$'s out of $k+2$ superfields. 

Let us now count the number of configurations of the first type
namely,
$\;$
${\rm Tr } ( (W^2)^{2p +1} \Phi^{k-2p} )$. This is obtained by
subtracting  the total number of configurations in the gauge
function proportional to $(W^2)^{2p +1}$ by $N_o$. The total number of
such configurations is given by $N_t = {}^{k+2} C_{2} {}^k C_{2p}$. 

Therefore the terms proportional to $g_k {W^2}^{2p+1}$ from both the
superpotential and gauge function totally are given by
\bea{termk}
-\frac{i}{4 \Lambda^2} g_k  \frac{  \alpha^{2p} }{(k+1)(k+2)} 
\left( ( N_t -N_o ) {\rm Tr} ( W^2)^{2p+1} \Phi^{k-2p} +  \right. \\
\nonumber
\left. N_o{\rm Tr} ( W^2 W^\alpha W^2 W_\alpha (W^2)^{2(p-1)} \Phi^{k-2p} ) 
\right) \\ \nonumber
+ \tilde{m} \frac{g_k}{k+1} \alpha^{2p+1} N_s {\rm Tr}(( W^2)^{2p +1}
\Phi^{k-2p} ) 
\eea
Substituting the values of $N_t, N_o, N_s$ 
and $\alpha = \frac{i}{4\Lambda^2 \tilde{m}}$,
we see that the above term reduces to 
\be{termcom}
- \frac{2\alpha^{2p+1}N_0 }{(k+1)(k+2)} {\rm Tr} ( W^2 W^\alpha
  \{W_\alpha , W^2\} (W^2)^{2(p-1)} \Phi^{k-2p} )
\ee
This is because the combinatorial factors satisfy the relation
\be{relcom}
N_0 = \frac{1}{4} \left( 2N_t - (k+2) N_s \right)
\ee
Now from the second equation of \eq{iden1} we see that the term in
\eq{termcom} is already proportional to $G^2$ or $G\cdot F$, therefore
reducing the remaining $W$'s by chiral ring identities ensures that
such terms vanish.

What remains to compute is the term proportional to $(W^2)^2$ coming
from the superpotential and the gauge function. From the
superpotential this term is 
\begin{equation}
\tilde{m} \alpha^2  \frac{g_k}{k+1} {}^{k+1}C_2  {\rm Tr}(\Phi^{k-1}(W^2)^2)
\label{w4sp}
\end{equation}
while from the gauge function, since there is no difference between
${\rm Tr}(\Phi^{k-1}(W^2)^2)$ and ${\rm Tr}(\Phi^{k-1}W^{\alpha} W^2 
W_{\alpha})$ due to the cyclicity of trace, we obtain
\begin{equation}
-3\alpha \frac{i}{4\Lambda^2}  \frac{g_k}{(k+1)(k+2)} {}^{k+2}C_3 {\rm Tr}
(\Phi^{k-1}(W^2)^2)
\label{w4gf}
\end{equation}
where the factor $3$ in the front appears due the three possible 
arrangements of $W^2$ and the two single $W_{\alpha}$'s.   
Substituting the value of $\alpha$ we find that the two contributions
(\ref{w4sp}) and (\ref{w4gf}) exactly cancel.

To conclude, we have shown here that after the field redefinition of
$\Phi$, so that the new field commutes with $W_{\alpha}$, all the
non-trivial $\Phi$- dependent (and hence $\Lambda$-dependent) part of
the gauge function cancels and we are left with just the $\Phi$ and
$\Lambda$ independent gauge function. This was the situation
considered
in \cite{Alday:2003ms}.

\bibliographystyle{utphys}
\bibliography{susbrek}

\end{document}